\pdfoutput=1

\documentclass[12pt,letterpaper]{article}
\usepackage{jheppub}
\usepackage{journals}
\usepackage{graphicx}
\usepackage{amssymb}
\usepackage{placeins}
\usepackage{jheppub}
\usepackage{journals}
\usepackage{graphicx}
\usepackage{amsmath}
\usepackage{placeins}
\usepackage{wrapfig}
\usepackage{subfig}
\usepackage{amssymb}
\usepackage{hyperref}
\usepackage[toc,page]{appendix}
\usepackage{tikz}

\begin		{document}

\title		{Collision of localized shocks in AdS$_5$  as a series expansion in transverse gradients}

\author[a]{Sebastian Waeber,}
\author[a]{Laurence G.~Yaffe}

\affiliation[a] {Department of Physics, University of Washington, Seattle WA 98195-1560, USA
}
\emailAdd	{swaebe@uw.edu}
\emailAdd	{yaffe@phys.washington.edu}

\abstract
{We introduce a computational framework to more efficiently calculate the collision of localized shocks in five dimensional asymptotically Anti-de Sitter space. We expand the Einstein equations in transverse gradients and find that our numerical results agree well with  exact solutions already at first order in the expansion. Moreover, the Einstein equations at first order in transverse gradients can be decoupled into two sets of differential equations. The bulk fields of one of these sets has only a negligible contribution to boundary observables, such that the computation on each time slice can be simplified to the solution of several planar shockwave equations plus four further differential equations for each transverse plane `pixel'. At the cost of  errors of $\lesssim 10 \%$ at the hydrodynamization time and for low to mid rapidities, useful numerical solutions can be sped up by roughly one order of magnitude.}
\maketitle
\section{Introduction}
\label{intro}
In the past decade a considerable  progress has been achieved with respect  to simulating heavy ion collisions via holography \cite{che3, Chesler:2010bi, Che, Chesler:2015fpa, wae3, mue, Casalderrey-Solana:2013aba, 1307.2539, 1507.08195, 1607.05273, 1609.03676, 1312.2956}. The difficulty of simulating the first few fractions of $1\, \text{fm}/c$ after a heavy ion collision comes from the strongly coupled and far from equilibrium nature of the incipient quark-gluon plasma. The gauge/gravity duality, or the  dynamical equivalence of strongly coupled, conformal, supersymmetric field theories in $d$ dimensions to weakly coupled Einstein gravity in $d+1$ dimensions, provides a powerful tool for studying the early stages of a heavy ion collision. The duality between gravity in five dimensional asymptotically Anti-de Sitter  (AdS) space and $\mathcal{N}=4$ super-Yang-Mills (SYM) theory is especially useful in this context, as the latter theory provides a reasonable approximation to a QCD plasma at high temperatures. \\ \indent Holographic simulations of heavy ion collisions have become progressively  more intricate and realistic up to the point of the exact treatment of  collisions of localized nuclei in \cite{Che}. These studies revealed non-trivial and sometimes unexpected properties of the pre-hydrodynamic quark gluon plasma, from a sizable transverse flow \cite{Che}, to  universal behaviors regarding post collision flow and hydrodynamization at near constant proper time \cite{che3, Chesler:2015fpa, wae3}, just to name a few. However, there are still many  interesting and demanding problems that remain unexplored. In particular the simulation of localized heavy ion collisions with a granular structure needed to explain the observed large event-by-event fluctuations of flow observables. In this work we develop tools to substantially facilitate the treatment of these problems by introducing a computational framework to efficiently calculate localized shockwave collisions in five dimensional asymptotically AdS space. \\
\indent Heavy ions collided at LHC and RHIC are highly relativistic and Lorentz contracted. Due to this strong contraction, gradients transverse to the beam axis are very small compared to longitudinal gradients. The idea of approximating heavy ion collisions by neglecting transverse gradients is not new and is  the justification for earlier studies of  planar shockwave collisions in Anti-de Sitter space \cite{che3, Chesler:2010bi}. Planar collisions can be thought of as the zeroth order solutions to Einstein equations that were expanded in transverse gradients. \\ \indent In this work we study the next-to-leading order results when the Einstein equations are expanded in transverse derivatives. Already at first order in transverse derivatives we find remarkably good agreement with the exact results in \cite{Che}. Moreover, we show that at first order the Einstein equations can be decoupled into two sets of differential equations for two disjoint sets of bulk fields. The fields of the larger set contribute negligibly to boundary observables. We  explain why this happens and find that we can reproduce the exact results to good accuracy for times  up to the hydrodynamization time, even for broad shocks, with an algorithm that only requires the planar shock solution and four additional differential equations to be solved on each time slice and transverse `pixel'. Just using a Mathematica implementation, we can reproduce, to good accuracy, the results of \cite{Che}, using 12 cores on a standard desktop machine, in about $36$ hours. \\ \indent The paper is structured as follows: For those readers less interested in computational details, we present key  results at the beginning, in section \ref{res_0}.  This is followed by a broader discussion of the transverse gradient expansion in section \ref{sec_transverse_gradient_expansion} and  more detailed discussion of our calculations in section \ref{shock_collisions}. Section \ref{res} contains  further analysis of our numerical results, followed by a few concluding remarks in section \ref{conclusion}.
\section{Approximating transverse flow and hydrodynamization time in heavy ion collisions}
\label{res_0}
\begin{figure}
\begin{center}
\includegraphics[scale=0.57]{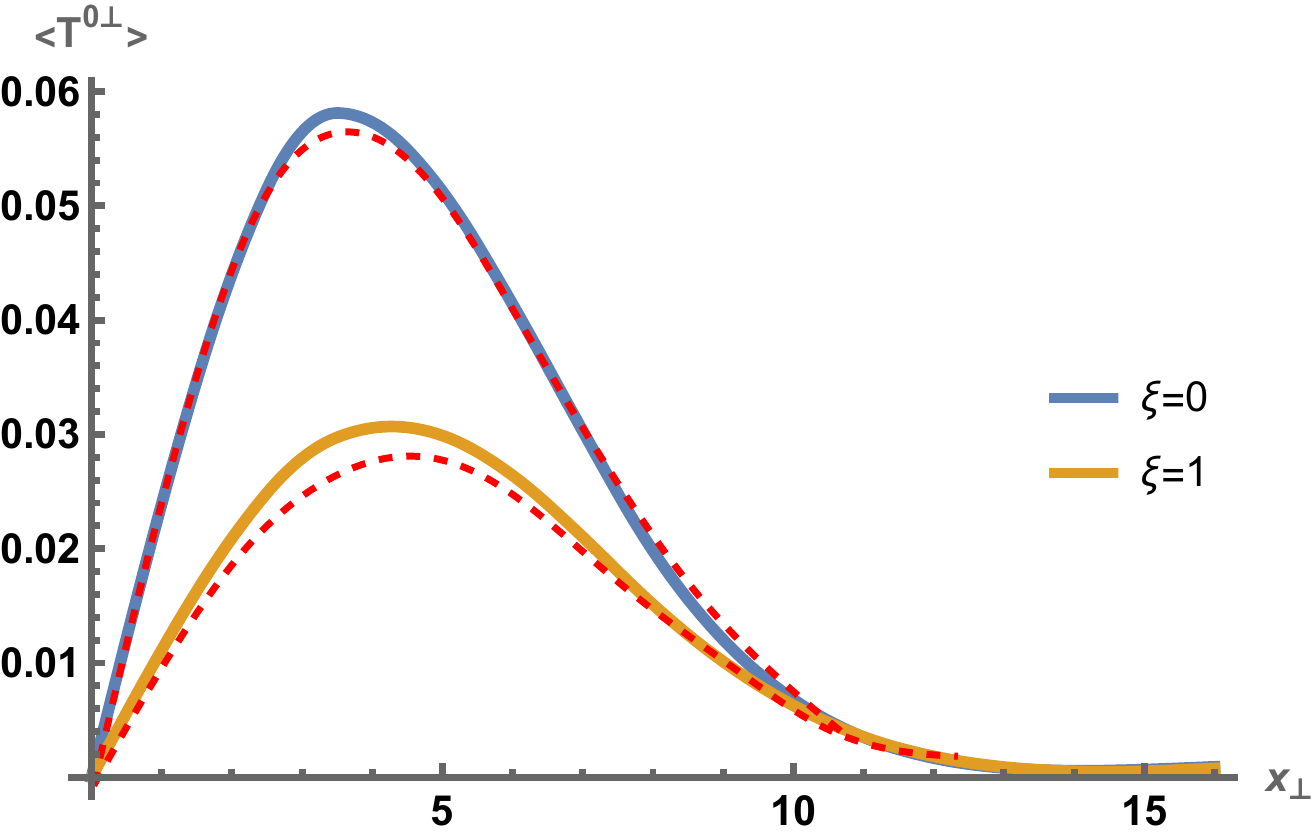}
\includegraphics[scale=0.57]{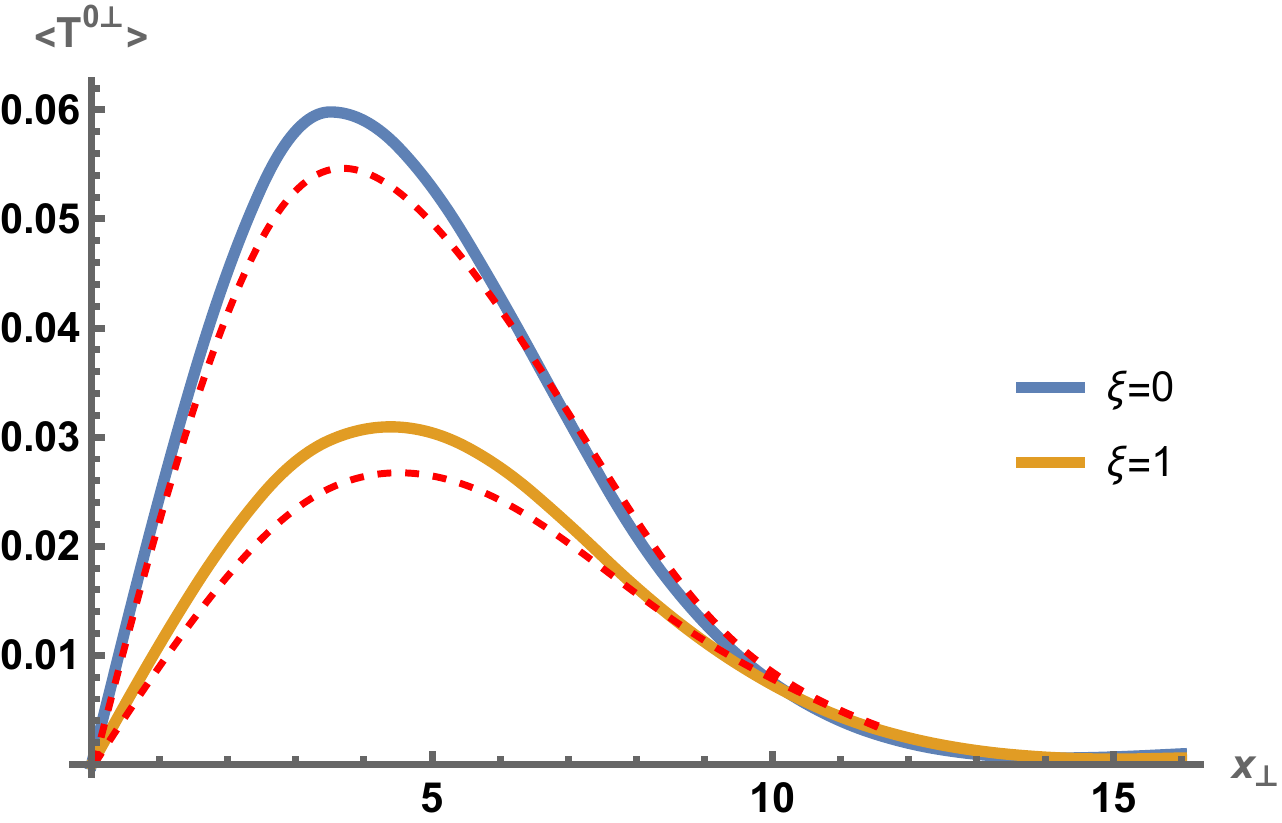}
\end{center}
\caption{The averaged, lab-frame  transverse momentum density $\langle T^{0 \bot}\rangle= \langle (\hat{x}_{\bot})^i \, T^{0i}\rangle$ up to first order  in transverse derivatives, as a function of the transverse plane radius $x_\bot= \sqrt{x^2+y^2}$, at proper time $\tau=1.25$ (left) and $\tau=2$ (right). Both plots display the averaged transverse energy flux at rapidity $\xi=0$ (blue curve) and at $\xi = 1$ (yellow curve). The red dashed curves represent the exact results of \cite{Che}. At $\tau=1.25$ and low rapidities  the first order approximation and the exact results agree very well. This agreement slightly deteriorates for larger rapidities and later proper times, but still remains quite good. We work in units in which the maximum of the longitudinally integrated energy density is set to 1.}
\label{transv}
\end{figure}

\begin{figure}
\begin{center}
\includegraphics[scale=0.75]{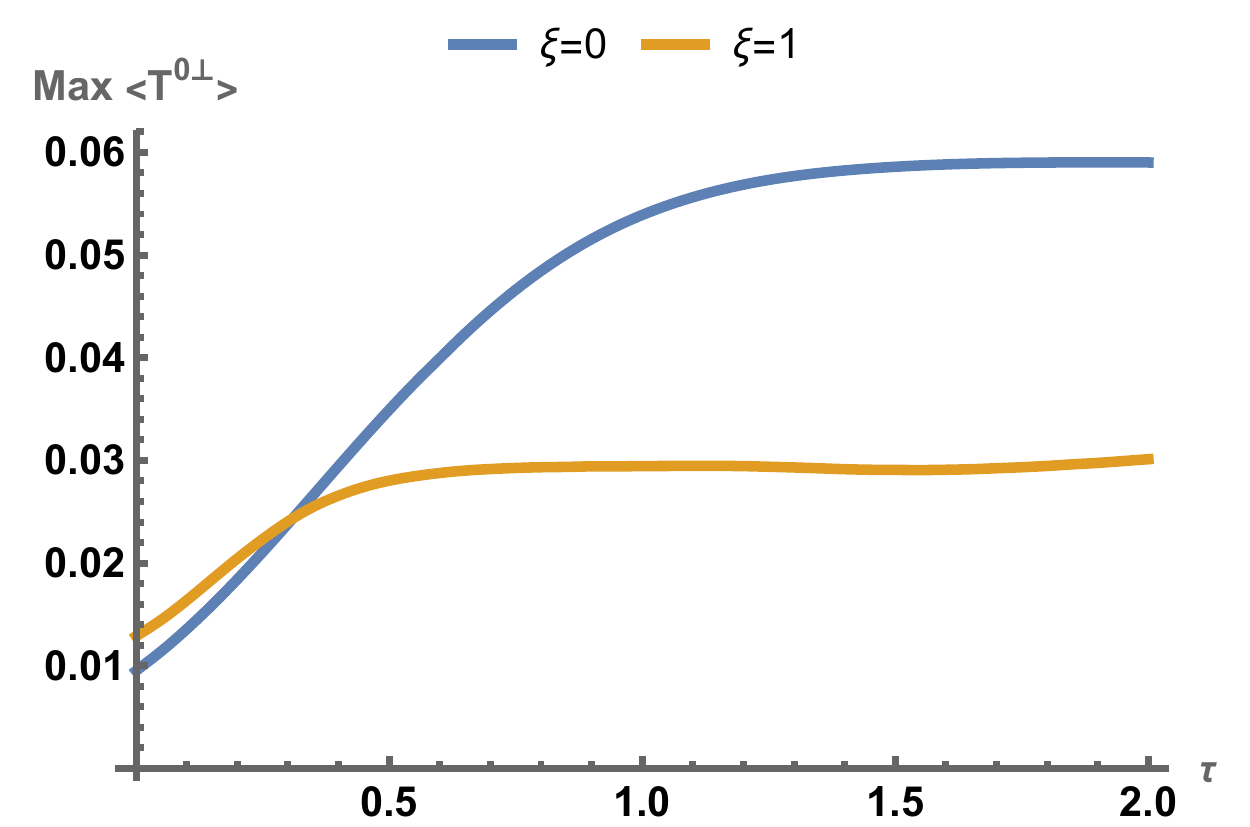}
\end{center}
\caption{The maximum value of the averaged, lab-frame  transverse momentum density $\langle T^{0 \bot}\rangle= \langle (\hat{x}_{\bot})^i \, T^{0i}\rangle$ up to first order  in transverse derivatives, as a function of proper time $\tau$,  for two rapidities $\xi=0$ (blue curve) and $\xi=1$ (yellow curve). The energy flux appears to reach a stable plateau, consistent with the approximation in \cite{Pratt}.}
\label{transvmax}
\end{figure}

\begin{figure}
\begin{center}
\includegraphics[scale=0.55]{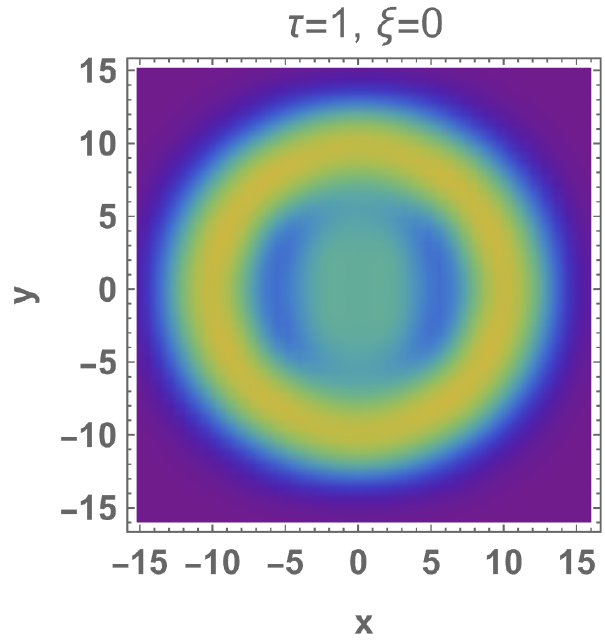}
\includegraphics[scale=0.55]{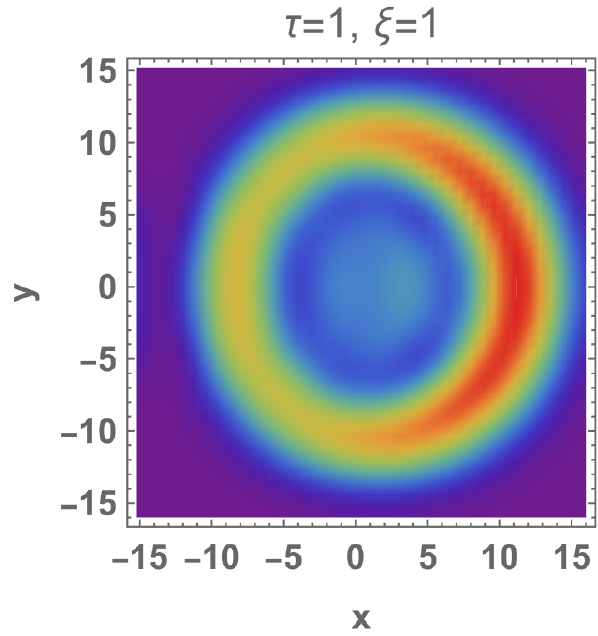}
\includegraphics[scale=0.55]{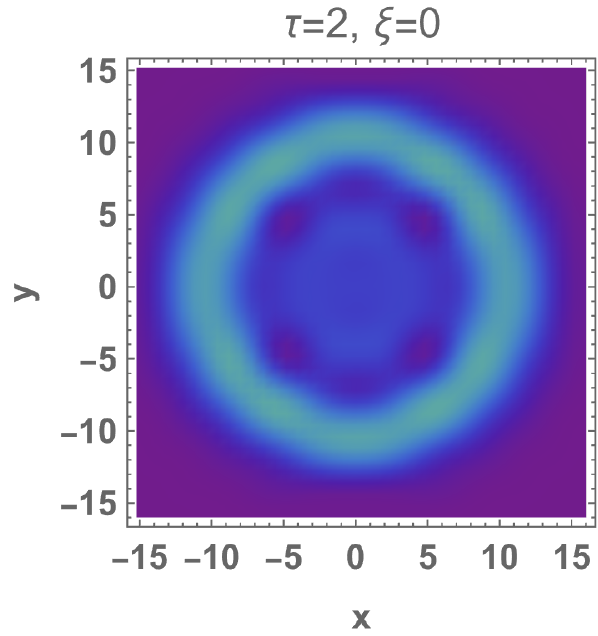}
\includegraphics[scale=0.55]{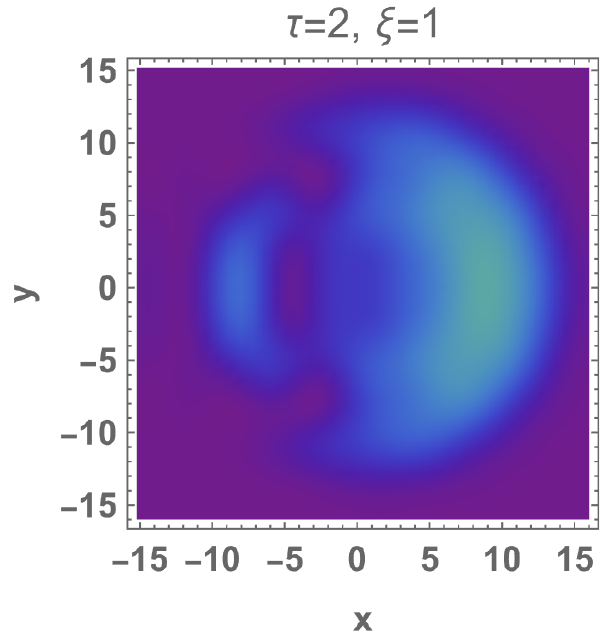}
\includegraphics[scale=0.4]{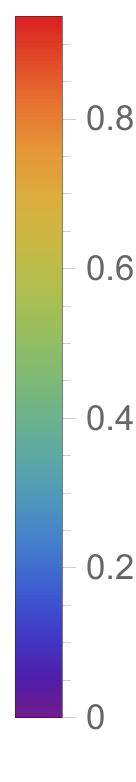}
\\
\includegraphics[scale=0.55]{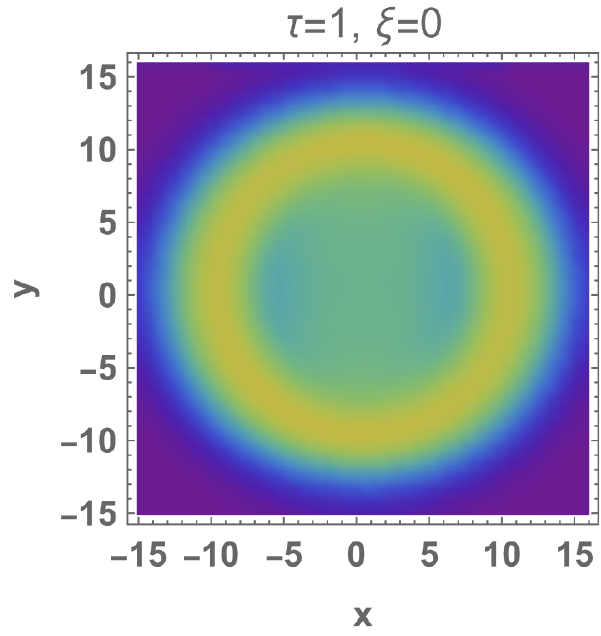}
\includegraphics[scale=0.55]{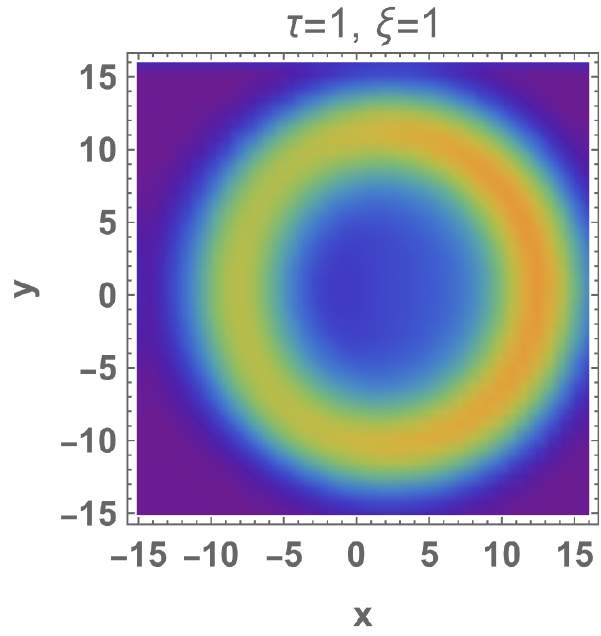}
\includegraphics[scale=0.55]{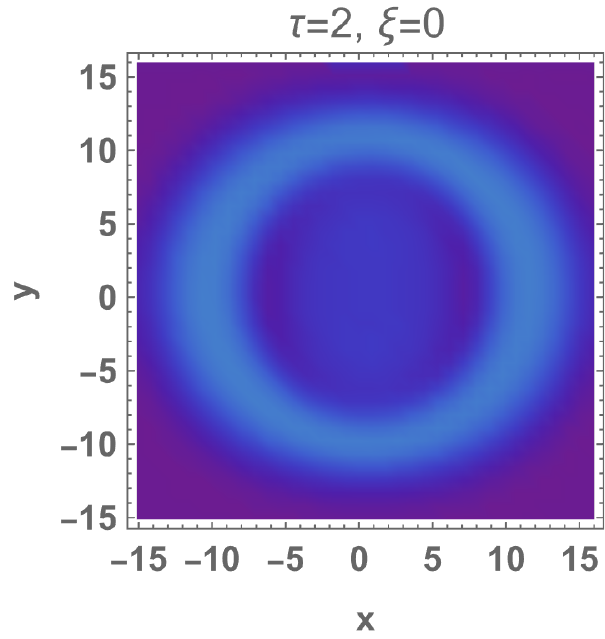}
\includegraphics[scale=0.55]{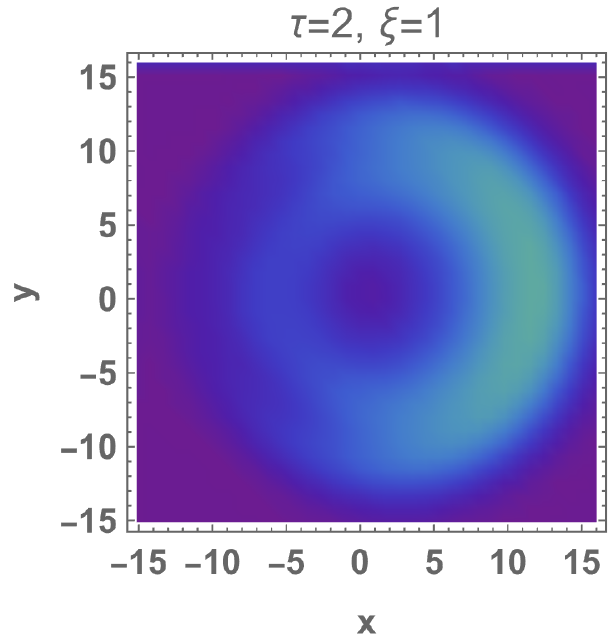}
\includegraphics[scale=0.4]{sc_comp.pdf}
\end{center}
\caption{The residual $\Delta$ (defined in Eq. (\ref{residual}))  indicating when and where the hydrodynamic approximation is a reasonable description. For $\Delta \lesssim 0.15$ hydrodynamics is applicable.   We show the results  for the zeroth order in transverse derivative approximation (first row) and the exact solutions (second row) from \cite{Che}, for collisions with otherwise identical parameters. Especially for low rapidities and close to the central region we see quite good agreement between the  exact and approximate solution, already at zeroth order in transverse derivatives. In both cases one finds a hydrodynamization time of $t \approx 1.25$ in the central region and for zero rapidity. Visible artifacts are a consequence of the sparse grids used to discretize the transverse directions.}
\label{residual_comparison}
\end{figure}

\iffalse In this section we are going to anticipate the arguably most interesting result of this paper. A more thorough analysis of our results can be found in section \ref{res}. \\ \indent \fi From \cite{Che} we know that during the early, pre-hydrodynamic stage after the collision of two relativistic heavy ions, a sizable transverse flow is produced, which simulations at zeroth order in transverse derivatives, e.g. \cite{che3, Chesler:2010bi, Che, Chesler:2015fpa, wae3, mue, Casalderrey-Solana:2013aba, 1307.2539, 1507.08195, 1607.05273, 1609.03676, 1312.2956}, fail to capture. While there are models, e.g. \cite{Pratt}, which roughly estimate the size of transverse momenta from  zeroth order data, the simplifications and  assumptions required do not yield  quantitative control or reliable error estimates.\\ \indent  The systematic approximation discussed in this work  requires the aspect ratios of projectiles to be large, an assumption which is very well fulfilled for realistic heavy ion collisions. We find that already for localized collisions of projectiles with a comparatively small aspect ratio of $8$, the transverse flow of our approximation and the exact results in \cite{Che} agree surprisingly well, both at mid and small rapidity, up until the hydrodynamization proper time. Examples of our results for transverse energy flux are depicted in  Fig. \ref{transv}. The error of the  first order approximation  increases slowly for larger rapidities and later proper times.  As suggested by the approximation in \cite{Pratt} and consistent with \cite{Che} we find that the maximum value of the transverse energy flux as a function of proper time reaches a plateau as displayed in Fig. \ref{transvmax} and stays approximately constant. For higher rapidities the plateau is reached at later proper times. 
\\ \indent 
One key observable to extract from holographic heavy ion collisions is the hydrodynamization time. To quantify how well the system is described by hydrodynamics one introduces a residual  $\Delta$ (defined later on in Eq. (\ref{residual})), which measures the deviation of the stress energy tensor from its hydrodynamic approximation. For a small $\Delta$, typically below $0.15$, the hydrodynamic approximation is appropriate. Remarkably we find that our approximate solution and the exact solution of \cite{Che} for $\Delta$ match well already at zeroth order in transverse derivatives, especially for low rapidities and in the central region of the collision. We also find approximately the same  hydrodynamization time  (i.e., the time at which the residual $\Delta$ drops below a threshold of $0.15$)  in the central region. We display the residual $\Delta$ and its zeroth order in transverse derivatives approximation in Fig. \ref{residual_comparison} at various proper times and rapidities. \\
\indent In general, observables evaluated before the hydrodynamization time are well described by either the zeroth or first order  transverse derivative approximation.

\section{Transverse gradient expansion}
\label{sec_transverse_gradient_expansion}
Let $G_{\mu \nu}$ be a bulk solution to Einstein equations in asymptotically  AdS space which varies slowly with respect to the transverse boundary coordinates $\bold x^\bot= (x,y)$ compared to its variation with respect to the boundary coordinate $x^{||}=z$. \iffalse Derivatives along $x_{||}$ in general do not have to be small, i.e. we do not assume that the hydrodynamic evolution of the boundary stress energy tensor is everywhere a valid approximation to the evolution of the stress energy tensor  computed from the bulk metric $G_{\mu \nu}$.\fi We introduce a formal parameter $\epsilon$  such that, after defining $ \tilde{\bold x}^\bot \equiv \epsilon\,  \bold x^\bot$, gradients along $\tilde{\bold x}^\bot$ and $x^{||}$ are of comparable order
\begin{equation}
\Big(\partial_{\tilde{\bold x}^\bot} G_{\mu \nu} \Big) / \Big(\partial_{x^{||}} G_{\mu \nu} \Big)=\mathcal{O}(1).
\end{equation}
 Since the metric $G_{\mu \nu}$ is assumed to only weakly depend on $\bold x^\bot$ the parameter $\epsilon$ is a small number and we can treat the Einstein equations written in terms of the rescaled coordinate $ \tilde{\bold x}^\bot$ as a perturbative expansion in $\epsilon$. We express the metric in terms of the rescaled coordinate $G_{\mu \nu}(x^0,x^{||},\bold x^{\bot}) = \tilde{G}_{\mu \nu}(x^0,x^{||},\bold {\tilde{x}}^{\bot})$ expand the Einstein equations in powers of $\epsilon$, truncate at a given order,  and then scale back to the original coordinates $ \bold x^\bot$ when solving the equations. \iffalse  For this we only expand the differential operators in $\epsilon$.  In principle the metric expressed in the new coordinate $\tilde{\bold x}_\bot$ does also depend on $\epsilon$ after the rescaling  $\bold x_\bot \rightarrow \tilde{\bold x}_\bot= \epsilon\,  \bold x_\bot$. However, we pretend that the $\epsilon$ parameter appearing in the metric and the $\epsilon$ parameter appearing in differential operators $\partial_{\bold x_\bot}=\epsilon \cdot \partial_{\tilde{\bold x}_\bot}$ are two distinct small numbers and only expand in the latter, while treating all functions exactly in the former. \fi  This procedure is equivalent to replacing $\partial_{\bot} \rightarrow \epsilon \,  \partial_{\bot}$, expanding the Einstein equations in powers of $\epsilon$, solving them order by order  and setting $\epsilon =1$ in the end. \\ \indent To spell this out more explicitly, we write the Einstein equations for a metric $G$ schematically as 
\begin{equation}
E(G)=0.
\end{equation} 
Expanding in transverse derivatives, we have
\begin{equation}
E(G)=E^{(0)}(G)+ \epsilon \, E^{(1)}(G)+ \epsilon^2\, E^{(2)}(G),
\end{equation}
where the differential operator $E^{(i)}$ contains $i$ powers of transverse derivatives.
   Let $G^{(i)}_{\mu \nu}$ denote an approximate solution to the  Einstein equations  valid to order $\mathcal{O}(\epsilon^i)$ so that \begin{equation}
E(G^{(i)})=\mathcal{O}(\epsilon^{i+1}).
\label{approximate_EQ}
\end{equation} We assume that $G^{(i)}_{\mu \nu}$ is known  up to some maximum order $I$ on some initial Cauchy surface at time $t=t_0$.  We want to solve for $G^{(i)}_{\mu \nu}$ up to this order $I$ for  some future time interval  $t_0\leq t \leq t_1$. At order zero the metric $G^{(0)}_{\mu \nu}(x^0,x^{||},\bold x^\bot)$ is, for every fixed value of $\bold x^\bot$, some solution to the planar Einstein equations (obtained by neglecting transverse derivatives), with parameters of the specific planar solution varying slowly with $x^\bot$. In other words
   \begin{equation}
   E^{(0)}(G^{(0)})=0.
   \label{lowest_order_EQ}
\end{equation}    
We emphasize that $G^{(0)}$ is not a single planar solution with vanishing transverse derivatives, rather the relevant planar solution varies (slowly) with changing $\bold x^\bot$.
One now systematically corrects this zeroth order approximation by writing
\begin{equation}
G^{(i)}_{\mu \nu}(x^0,x^{||},\bold x^\bot)=G^{(i-1)}_{\mu \nu}(x^0,x^{||},\bold x^\bot)+\delta g_{\mu \nu}^{(i)}(x^0,x^{||},\bold x^\bot)
\end{equation} 
 and demands that the Einstein equations hold up to the next order.
\iffalse 
We expand $E(G)$ in the following way
\begin{equation}
E(G)=E^{(0)}(G)+E^{(1)}(G)+E^{(2)}(G),
\end{equation}
where the differential operator $E^{(i)}$ contains $i$ powers of transverse derivatives. \fi Let $\Delta_L^{(i)}$ be the planar Lichnerowicz operator  evaluated on $G^{(i)}$,
\begin{equation}
\Delta_L^{(i)}  \equiv\frac{ \delta E^{(0)} (G^{(i)})}{\delta G^{(i)}}.
\end{equation}
Then Eq. (\ref{approximate_EQ}) will be satisfied if
\begin{equation}
\Delta_L^{(i-1)}  \delta g^{(i)} = -\epsilon \,E^{(1)}(G^{(i-1)}) -\epsilon^2\,E^{(2)}(G^{(i-2)}).
\label{expanded_EQ}
\end{equation}
  \section{Transverse derivative corrections to shock collisions}
  In the next two subsections we  quickly  review the general treatment of (localized) shock collisions in AdS$_5$ \cite{che3, Chesler:2010bi, Che}, after which we discuss in more detail the solution of the Einstein equations in the transverse gradient expansion up to first order. We address how to construct consistent initial data up to first order in transverse derivatives in section \ref{initial_data}. \iffalse and finally identify those contributions to second order in transverse gradients, that can be effortlessly computed from solution that solve Einstein equations up to first order in transverse gradients in subsection \ref{partial}.\fi
  \label{shock_collisions}
  \subsection{Single localized shocks in Fefferman-Graham coordinates}
  \label{single_shock}
  
  \iffalse  Colliding projectiles with non-trivial transverse dependence and ignoring derivatives in transverse direction appearing in the Einstein equations is the same as running multiple planar shockwave collisions in parallel, with the initial shock widths and amplitudes being parametrized by $\bold x^\bot$. \fi  As in \cite{che3, Che},  a  single shock moving in $\pm z$ direction in Fefferman-Graham (FG) coordinates may be described  analytically,
\begin{equation}
ds^2= \frac{1}{\rho^2}\big(-dt^2+d\rho^2+(d\bold x^\bot)^2+dz^2+\rho^4 h_{\pm}(\bold x^\bot,z^\mp,\rho) (dz^\pm)^2 \big)
\label{FG}
\end{equation}
with $z^\mp = z \mp t$. The metric ansatz (\ref{FG}) solves the Einstein equations  provided the function $h_\pm$ fulfills the linear differential equation
\begin{equation}
\Big(\frac{d^2}{d \rho^2}  -\frac{3}{\rho} \frac{d}{d \rho} +\,\nabla_\bot^2\Big) \rho^4\, h_\pm =0.
\label{Ein_init}
\end{equation}
Any $\rho$-independent function, $h_\pm(\bold x^\bot,z^\mp,\rho)=h_\pm(\bold x^\bot,z^\mp)$, solves (\ref{Ein_init}) through first order in transverse derivatives. To match previous work, we will use a Gaussian profile\iffalse 
 In this work we focus on corrections to planar shockwaves up to first order in transverse derivatives and 
up to  this order the ansatz \fi
\begin{equation}
h_\pm(\bold x^\bot,z^\mp)= \frac{\mathcal{A}}{\sqrt{2 \pi w^2}} \exp\big(-\frac{1}{2}(z^\mp)^2/w^2 \big)\exp \big( -\frac{1}{2} (\bold x^\bot -\bold b)^2/R^2 \big).
\label{hpm}
\end{equation} With this choice, the single shock metric is a valid zeroth order solution solving (\ref{lowest_order_EQ}), which we denote as 
 $G^{(0)}[h_\pm]$. Moreover the same metric is a valid solution at first order, so  $G^{(1)}_{FG}[h_\pm] = G^{(0)}_{FG}[h_\pm]$.  In general the  metric (\ref{FG}) corresponds to a state in the dual field theory for which
\begin{align}
\langle T^{00} \rangle &=\langle T^{zz} \rangle = \frac{N_c^2}{2 \pi^2} h_{\pm}\Big|_{\rho=0}\\
\langle T^{0z} \rangle &= \pm \frac{N_c^2}{2 \pi^2} h_{\pm}\Big|_{\rho=0}
\end{align}
\iffalse To compute initial conditions for the collision of shocks up to order $\mathcal{O}(\nabla_{\bot})$ we add up two  single shock solutions that are spatially well separated on an initial spacelike surface. In the next two sections we will address the integration strategy and then discuss in more detail how we transform the initial data for (spatially well separated) left and right moving single shock solutions from Fefferman-Graham coordinates to Eddington-Finkelstein coordinates, a coordinate system that allows a numerically efficient integration forward in time. After the transformation we will add the well separated left and right moving single shock solution to construct the initial data.
\fi
\iffalse
  The metric ansatz  (\ref{FG}) with (\ref{hpm}) evaluated on an initial spacelike surface  will serve as initial conditions for the evolution of single shocks. To compute collisions of shocks up to order $\mathcal{O}(\nabla_{\bot})$ we add up two  single shock solutions that are spatially well separated on the initial surface.
  \fi
\subsection{Integration strategy}
\label{Integration_strategy}
For the time evolution of two colliding localized shocks, we transform to infalling Eddington-Finkelstein (EF) coordinates and employ the characteristic formulation of general relativity. In EF coordinates the line element of the metric $G_{\mu \nu}$ may be written as
 \begin{equation}
 ds^2 = r^2 h_{\mu \nu}(x,r)dx^\mu dx^\nu +2\, dr dt.
 \label{EF}
 \end{equation}
 Here $x \equiv (t,\bold x^\bot, x^{||})$.
 In these coordinates the boundary is positioned at $r = \infty$.
 Following \cite{che3}, we have used  diffeomorphism invariance to fix the component of the line element proportional to  $dr dt$ to equal  $2\, dr dt$.  The metric (\ref{EF}) is also invariant under shifts of the radial coordinate of the form
 \begin{equation}
 r \rightarrow r'= r+ \lambda(x).
 \label{shift_parameter}
 \end{equation}
 As in  \cite{che3} we will exploit the radial shift symmetry to fix the radial position of the apparent horizon $r_h$ to correspond to the end of our integration domain. \\ \indent  Let us write the metric in EF coordinates more specifically as 
  \begin{equation}
 ds^2 = -2 A(x,r)dt^2+G_{ij}(x,r)dx^i dx^j+2 dr dt-2 F_i(x,r)dt dx^i,
 \label{EF_2}
 \end{equation}
 with $i=x,y,z$ denoting spatial derivatives.
 Let $G_{ij} =\Sigma^2 \hat{g}_{ij}$ with $\det \hat{g}_{ij}=1$. Note that, unlike in section \ref{sec_transverse_gradient_expansion}, henceforth $G_{ij}$ denotes the spatial part of the EF-metric. For convenience we also slightly modify our convention regarding the superscript index $(\cdot)^{(i)}$. Henceforth, for a  function $X$, let $X^{(i)}$ denote the difference between the  $i$-th and $(i-1)$-th order solution. In other words, using the notation of section \ref{sec_transverse_gradient_expansion}, we abbreviate $\delta X^{(i)}$ with $X^{(i)}$ and we will henceforth use the notation $X^{(i)}$ only in this context.  Through  first order in transverse derivative expansion the unimodular condition on $\hat g$ implies that $\det \hat{g}^{(0)}_{ij}=1$, and $\text{Tr}\,\hat{g}^{(1)}_{ij}\equiv (\hat g^{(0)})^{ij}(\hat g^{(1)})_{ij}=0$ for the first order correction in transverse derivatives $\hat{g}^{(1)}$. The spatial scalar factor $\Sigma$ has the near boundary form \begin{equation}
 \Sigma = (r+\lambda)+\mathcal{O}(r^{-7}),
\end{equation} and the other metric functions have a near boundary beahvior given by  
\begin{align}
\label{expand_nb_1}
(r+\lambda)^{-2}\,A &= \frac{1}{2}+r^{-4}\, a^4+\mathcal{O}(r^{-5}),\\
(r+\lambda)^{-2}\,F_i &= -\frac{\partial_i \lambda}{(r+\lambda)^2}+r^{-4}\, f_i^4+\mathcal{O}(r^{-5}), \\
(r+\lambda)^{-2}\,\hat{g}_{ij} &= \frac{\delta_{ij}}{(r+\lambda)^2}+r^{-4}\,\hat{g}^{4}_{ij}+\mathcal{O}(r^{-5}).
\label{expand_nb_3}
\end{align} 
Let $a^{4}$, $f_i^4$ and $\hat{g}^{4}_{ij}$  be the $\mathcal{O}(r^{-4})$ expansion coefficients of the near boundary expansion of the rescaled functions $(r+\lambda)^{-2}\,A$, $(r+\lambda)^{-2}\,F_i$ and $(r+\lambda)^{-2}\,\hat{g}_{ij}$.
  The CFT stress energy tensor in $3+1$ dimensions  is determined by the $\mathcal{O}(r^{-4})$ coefficients $a^{4}$, $f_i^4$ and $\hat{g}^{4}_{ij}$ in the near boundary expansions (\ref{expand_nb_1})-(\ref{expand_nb_3}):
 \begin{align}
 \langle T^{00}\rangle& = -\frac{3}{2} \big ( (a^4)^{(0)}+ (a^4)^{(1)}\big)+\mathcal{O}(\epsilon^2) \\
  \langle T^{0z}\rangle &= - \big ( (f_z^4)^{(0)}+ (f_z^4)^{(1)}\big)+\mathcal{O}(\epsilon^2) \\
    \langle T^{0\bot}\rangle &= -  (f_\bot^4)^{(1)}+\mathcal{O}(\epsilon^2) \\
 \langle T^{ij}\rangle& =  (\hat{g}_{ij}^4)^{(0)}+ (\hat{g}_{ij}^4)^{(1)}-\frac{1}{2}\big ( (a^4)^{(0)}+ (a^4)^{(1)}\big)\delta^{ij}+\mathcal{O}(\epsilon^2).
 \end{align}
 As demonstrated in \cite{Bondi:1960jsa, Sachs:1962wk} and generalized in \cite{che3} the Einstein equations expressed in EF coordinates (\ref{EF}) can be written as a nested system of ordinary radial differential equations (ODEs) on each time slice. Defining derivatives along outgoing geodesics, $d_+ = \partial_t + A\, \partial_r$, one can write these nested equations  in the form (\ref{E:SigmaEOM})-(\ref{E:AEOM}) given in  Appendix \ref{Einstein_equations}. 
  This system can be solved sequentially to determine the time derivatives of fields once we know $\{a^4,f_i^4,\hat{g}_{ij},\lambda\}$ on a given time slice. The shift $\lambda$ can be chosen to fix the horizon position equal to the endpoint of the integration domain in the bulk. \iffalse both  in order $\mathcal{O}(\nabla_{\bot}^0)$ and $\mathcal{O}(\nabla^1_{\bold x_\bot})$.\fi     
 Three of the differential equations are constraint equations, of which two are equivalent to the covariant conservation of the boundary stress energy tensor, which are used to compute the time derivatives of the energy and momentum density $\{a^4,f_i^4\}$ given by \begin{equation}
 \label{time_der_energy_(flux)}
\partial_t a^4 =\frac{2}{3} \partial_i f^4_i \,\,\,\,\,\,\,\,\, \partial_t 
f^4_i=\frac{1}{2}\partial_i a^4-\partial_j \hat g^4_{ij}.
\end{equation} \iffalse  The time derivative of $\hat{g}_{ij}$ can be determined after the system of ODEs are solved which determine $A$ and $d_+\hat g$. The time derivative is then given by
\begin{equation}
 \label{time_der_metric}
\partial_t \hat g_{ij}= d_+ \hat g_{ij}-A\,\partial_r \hat g_{ij}.
\end{equation}
\fi

  The differential equation that determines the shift $\lambda$, so that the horizon matches the end of our integration domain, is obtained by requiring that the congruence of outgoing geodesics has vanishing expansion rate there. This results into the non-linear ODE
\begin{equation}
\label{Eq_lambda}
d_+\Sigma|_{r_h}= -\frac{\Sigma' F^2}{2}-\frac{\Sigma \, \nabla F}{3}\Big|_{r_h},
\end{equation}  
where $r_h$ is the radial position of the apparent horizon, primes $'$ denote radial derivatives $\partial_r$ and $ \nabla$ is the covariant derivative acting on spatial tensor fields  using the Christoffel connection of $G_{ij}$ (see (\ref{Gamma})).
   Requiring this expansion rate  to be constant at the radial position of the horizon leads to a linear elliptic differential equation for $\partial_t \lambda$.  Having solved for $\{\partial_t a^4,\partial_t f_i^4,\partial_t \hat{g}_{ij},\partial_t \lambda\}$ on one time slice, we use the fourth order Runge Kutta method to evolve the metric in time. \\ \indent In the case of exact localized shock collisions \cite{Che} the elliptic differential equation for $\partial_t \lambda$  is  difficult to handle and solving it requires a large amount  of wall clock time and memory\footnote{After discretization with $N_x$, $N_y$ and $N_z$ grid points in $x$, $y$ and $z$ directions, the standard brute force approach to solving this differential equation  would require the inversion of a matrix of size $(N_x \, N_y\, N_z)\times (N_x \, N_y\, N_z)$. For modest spatial grid sizes of, e.g., $32 \times 32 \times 512$ this requires over $2 $ TB of memory and more wall clock time than solving all other equations on a given time slice.  More sophisticated methods involving domain decomposition, require very large amounts of memory and run time as well.}.  The treatment of this equation is substantially simplified if we expand  in transverse derivatives. The details of this differential equation are  not relevant for the following discussion, for the interested reader, the exact differential equation is given in (3.47) of \cite{che3}. Let us 
write this linear horizon fixing equation schematically as
\begin{equation}
\Big[ K_{ij}\partial_i \partial_j\, +L_i \partial_i\, +M\Big]\,(\partial_t \lambda) -S=0,
\label{ell}
\end{equation}
where $K_{ij}$, $L_i$, $M$ and $S$ are functions of the metric and its derivatives.   At zeroth order in transverse derivatives  Eq. (\ref{ell})  simplifies to an  ordinary linear differential equation in the longitudinal coordinate $z$, which can be trivially  parallelized over the transverse coordinates $\bold x^\bot$,
\begin{align}
&D^{(0)} (\partial_t \lambda)^{(0)}=S^{(0)},\label{ell3}\\
&D^{(0)}=K^{(0)}_{zz}\partial_z^2+L_z^{(0)} \partial_z+M^{(0)}. \label{ell2}
\end{align}\iffalse
where we numerically compute the Green's function of (\ref{ell2}). \fi\iffalse This problem can be parallelized trivially over the transverse coordinates $\bold x^\bot$. \fi To compute the first order in transverse derivative correction we can reuse the Green's function for $D^{(0)} $ and transverse derivatives only contribute to the source term $\tilde S^{(1)}$,
\begin{align}
&D^{(0)} (\partial_t \lambda)^{(1)}=\tilde S^{(1)}\label{ell5}\\
\tilde S^{(1)}=S^{(1)}+\big(K^{(1)}_{zz}\partial_z \partial_z+L_z^{(1)} \partial_z+&M^{(1)}+\sum_{ij=x,y}K^{(0)}_{ij}\partial_i \partial_j+\sum_{i=x,y} L_i^{(0)} \partial_i \big)(\partial_t \lambda)^{(0)}. \label{ell4}
\end{align} Eq. (\ref{ell5}) is clearly still parallelizable over every value of $\bold x^\bot$. The ability to parallelize and  reuse the Green's function for $D^{(0)}$ allows us to circumvent this bottleneck of the exact   treatment. \\ \indent The same principle applies to the Einstein equations written as a system of ODEs that we solve on each time slice. Again we can reuse the radial Green's functions computed at zeroth order in transverse gradients.  The main  advantage of an expansion in transverse gradients is thus that Green's functions for each point in the transverse plain are no
more expensive to compute than in the planar case, both for the nested system of radial ODEs and the elliptic differential equation which fixes the horizon position. We solve the system of ODEs and (\ref{ell})-(\ref{ell4}) using spectral methods \cite{Boyd:Spectral}. The details of our numerics  can be found at the beginning of section \ref{res}.
\subsection{Initial data up to order $\mathcal{O}(\nabla_{\bot})$}
\label{initial_data}
We now discuss the construction of a numerical solution for the initial data of  localized shocks, in five dimensional AdS space, up to first order in transverse gradients $\nabla_{\bot}$. \iffalse 
For shocks with sufficient spatial separation the metric ansatz in FG coordinates, where we replace $h_{\pm}$ in (\ref{FG})  with $h_{+}+h_{-}$ solves the Einstein equations and serves together with (\ref{hpm}) as our initial data.  \fi
Following \cite{che3} we transform the  single shocks (\ref{FG}) from FG coordinates to infalling Eddington-Finkelstein (EF) coordinates.  In EF coordinates the metric takes the form
(\ref{EF}).
In this coordinate system the family of curves $\gamma_{x_0^\mu}(r)=(x_0^\mu,r)$  with varying radial coordinate $r$ is a geodesic congruence with affine parameter $r$.  Demanding that the tangential vector field $Y^\mu$ of the image of this congruence under the coordinate transformation to FG coordinates solves the geodesic equation, $Y^\mu\nabla_\mu Y^\nu=0$, provides a system of differential equations whose solution allows us to explicitly express FG  coordinates $Y(X)$ in terms of EF coordinates $X$. As boundary conditions of the geodesic equations we require that the coordinate systems match on the boundary, $\lim_{\rho \to 0}Y =\lim_{r \to \infty}X$, and that the radial coordinates match at a hypersurface in the bulk where $\rho_{max} = \frac{1}{r_{min}}$ with $r_{min}$  the end of our integration domain, $ r\in[r_{min},\infty]$.\\ \indent  We solve the geodesic equations up to first order in transverse derivatives, which contribute  via  derivatives acting on the metric in the Christoffel symbols\footnote{Even though it might appear so at first glance, writing the geodesic equation $Y^\mu\nabla_\mu Y^\nu=0$ in the form $\partial_\tau^2 x^\nu = -\Gamma_{\alpha \beta}^\nu \partial_\tau x^\alpha \partial_\tau x^\beta$ (with $Y^\nu = \partial_\tau x^\nu$) does not introduce an ambiguity regarding the transverse derivative expansion, since the relation $Y^\mu \partial\mu Y^\nu = \partial_\tau^2 x^\nu$ holds up to every order in $\epsilon$. Thus, the most convenient way to expand the geodesic equation in transverse derivatives is to expand the
equation $\partial_\tau^2 x^\nu = -\Gamma_{\alpha \beta}^\nu \partial_\tau x^\alpha \partial_\tau x^\beta$. It is then easy to see why transverse derivatives only explicitly contribute via the Christoffel symbols.}.   We write the zeroth order ansatz plus first order corrections for the FG coordinates $Y(X)$ in terms of the EF coordinates $X^\mu = (t, \bold x^\bot,z,u)$ with inverted radial coordinate $u=1/r$  as
\begin{align}
\label{coord_ansatz_s}
Y^\mu &= \big(t_{FG},( \bold x^\bot)_{FG},z_{FG},\rho\big),\\
t_{FG}&=t+u^2 \big(\alpha^{(0)}(X^\mu)+\alpha^{(1)}(X^\mu)\big),\\
( \bold x^\bot)_{FG}&= \bold x_\bot+u^5 \bold{\delta}_\bot^{(1)}(X^\mu),
\\
z_{FG}&=z+u^5 \big(\gamma^{(0)}(X^\mu)+\gamma^{(1)}(X^\mu)\big),
\\
\rho&=u+u^2 \big(\beta^{(0)}(X^\mu)+\beta^{(1)}(X^\mu)\big),
\label{coord_ansatz}
\end{align}
\iffalse where the superscript $^{(i)}$ denotes the contribution at the $i$-th order in transverse derivatives.\fi  With  (\ref{coord_ansatz_s})-(\ref{coord_ansatz}) we solve 
\begin{align}
& (Y^\mu)^{(0)}_{,uu}+\frac{2}{u} (Y^\mu)^{(0)}_{,u}+\big((\Gamma_{FG})^\mu_{\alpha \beta}\big)^{(0)} (Y^\alpha)^{(0)}_{,u} (Y^\beta)^{(0)}_{,u} =0,\\
& (Y^\mu)^{(1)}_{,uu}+\frac{2}{u} (Y^\mu)^{(1)}_{,u}+2\big((\Gamma_{FG})^\mu_{\alpha \beta}\big)^{(0)} (Y^\alpha)^{(1)}_{,u} (Y^\beta)^{(0)}_{,u} =-\big((\Gamma_{FG})^\mu_{\alpha \beta}\big)^{(1)}(Y^\alpha)^{(0)}_{,u} (Y^\beta)^{(0)}_{,u}.
\end{align} Here and henceforth we use the usual subscript notation for derivatives $(\cdot)_{,x}=\partial_x (\cdot)$ for a coordinate $x$ .  Let $G_{ab}$ denote the spatial part of the metric in EF coordinates and $\tilde{g}_{\alpha \beta}$ be the metric in FG coordinates. Then
\begin{equation}
 G^{(0)}_{ab}= 
\begin{pmatrix}
1\,\,\,\,\,\,\,\,\,\,\,\,\,\, & 0\,\,\, & 0 \\
0\,\,\,\,\,\,\,\,\,\,\,\,\,\, & 1\,\,\, & 0 \\
0\,\,\,\,\,\,\,\,\, \,\,\,\,\,& 0\,\,\, & (Y^\mu)^{(0)}_{,z} (Y^\nu)^{(0)}_{,z} \tilde{g}^{(0)}_{\mu \nu }
\end{pmatrix}
\label{metric_no1}
\end{equation}\iffalse
\begin{equation}
 G^{(1)}_{ab}= 
\begin{pmatrix}
1 & 0 & 2 (Y^\mu)^{(0)}_{,x}(Y^\nu)^{(0)}_{,z}\tilde{g}^{(0)}_{\mu \nu }  \\
0 & 1 & 2(Y^\mu)^{(0)}_{,y} (Y^\nu)^{(0)}_{,z} \tilde{g}^{(0)}_{\mu \nu } \\
2 (Y^\mu)^{(0)}_{,x} (Y^\nu)^{(0)}_{,z} \tilde{g}^{(0)}_{\mu \nu } & 2(Y^\mu)^{(0)}_{,y} (Y^\nu)^{(0)}_{,z} \tilde{g}^{(0)}_{\mu \nu } & \sum_{i+j+k=1}(Y^\mu)^{(i)}_{,z} (Y^\nu)^{(j)}_{,z} \tilde{g}^{(k)}_{\mu \nu }
\end{pmatrix}.
\label{metric_no2}
\end{equation}
\fi
and
\begin{align}
\label{metric_no2s}
 G^{(1)}_{ab} &= \delta_{ab}\,\,\,\,\,\,\,\,\,\,\,\,\,\,\,\,\,\,\,\,\,\,\,\,\,\,\,\,\,\,\,\,\,\,\,\,\,\,\,\,\,\,\,\,\,\,\,\,\,\,\,\,\,\,\,\,\,\,\,a,b=x,y\\
 G^{(1)}_{az} &=2(Y^\mu)^{(0)}_{,a} (Y^\nu)^{(0)}_{,z} \tilde{g}^{(0)}_{\mu \nu }\,\,\,\,\,\,\,\,\,\,\,\,\,\,\,\,\,\,\,\,\,\,\,\,a=x,y\\ 
 G^{(1)}_{zz} &= \sum_{i+j+k=1}(Y^\mu)^{(i)}_{,z} (Y^\nu)^{(j)}_{,z} \tilde{g}^{(k)}_{\mu \nu }.
\label{metric_no2}
\end{align}
For the numerical time evolution  we will need the zeroth order terms  $(\hat{g}_{ab})^{(0)}_\pm$ and first order corrections  $(\hat{g}_{ab})^{(1)}_\pm$ of the rescaled spatial metric $\hat g$. These are given by
\begin{align}
(\hat{g}_{ab})^{(0)}_\pm&=\big ( \det G^{(0)} \big)^{-1/3} G_{ab}^{(0)},\\
(\hat{g}_{ab})^{(1)}_\pm&=\big ( \det G^{(0)}  \big)^{-1/3} G_{ab}^{(1)}-\frac{1}{3} \big ( \det G^{(0)}  \big)^{-1/3} \,G^{(1)}_{ij} (G^{(0)})^{ij}  \,G_{ab}^{(0)},
\end{align}
such that $ \det( \hat g_\pm^{(0)}+\hat g_\pm^{(1)})=1+\mathcal{O}(\epsilon^2) $. The fourth order near boundary expansion coefficients of $A$ and $F_i$ in (\ref{EF_2}) for a single Gaussian shock moving in $\pm z$ direction are 
\begin{align}
(a^4)_\pm &= -\frac{2A}{3\sqrt{2 \pi w^2}} \exp\big(-\frac{1}{2}(z^\mp)^2/w^2 \big)\exp \big( -\frac{1}{2} (\bold x^\bot -\bold b)^2/R^2 \big) +\mathcal{O}(\epsilon^2)\\
(f_z^4)_\pm&=\frac{A}{\sqrt{2 \pi w^2}} \exp\big(-\frac{1}{2}(z^\mp)^2/w^2 \big)\exp \big( -\frac{1}{2} (\bold x^\bot -\bold b)^2/R^2 \big) +\mathcal{O}(\epsilon^2)\\
\bold f^4_\bot&=0.
\end{align}
Comparing the near boundary expansion of the determinant of the spatial metric given in (\ref{metric_no1}), (\ref{metric_no2s})-(\ref{metric_no2}) expressed by the functions $\alpha$, $\beta$, $\gamma$, $\delta$ and their derivatives, with the near boundary expansion of $\Sigma$, shows that the shift parameter $\lambda$, introduced in (\ref{shift_parameter}), can be obtained from the relation
\begin{equation}
(\lambda)^{(i)}_\pm=-\frac{1}{2}\big(\partial_u^2 Y^\rho \big)^{(i)}\big|_{u=0}.
\end{equation}
Finally the initial data for the collision of localized shocks at order $\mathcal{O}(\epsilon^i)$ in EF coordinates  at $t=t_0=-2$ is given by combining the data for well-separated counter-propagating single shocks,
\begin{align}
\{(a^4)^{(i)}_+ +(a^4)^{(i)}_-,\, \bold f_\bot^{(i)},\, (f_z^4)^{(i)}_+-(f_z^4)^{(i)}_-,\,(\hat{g}^{(i)}_{ab})_+ + (\hat{g}_{ab})_-,\,(\lambda)^{(i)}_+ +(\lambda)^{(i)}_-\}.
\end{align}
 Since each shock individually is an exact solution to Einstein equations, their sum is an approximate solution with exponentially small errors, if they are sufficiently separated, such that they don't overlap and the region between them is pure AdS up to exponentially small errors.
The initial time $t_0=-2$ is chosen such that left and right moving shocks are spatially well separated and have no overlap in the bulk region between the apparent horizon and the boundary. As in \cite{Che,Chesler:2010bi,che3}  we work with a small background energy density to deal with irregularities in the integration domain.
\subsection{Hierarchy among $\mathcal{O}(\nabla_{\bot})$-fields}
\label{hierarchy}
 The Einstein equations, expanded in transverse derivatives through first order, separate into two sets of differential equations for two disjoint sets  of bulk fields $S_1$ and $S_0$:
\begin{align}
S_1&=\{\,d_+ \hat g_{xz},\, d_+ \hat g_{yz},\,\hat g_{xz},\,\hat g_{yz},\, F_x,\, F_y\},
\label{set1}
\\
S_0&=\{\Sigma,\, d_+\Sigma,\,d_+ \hat g_{xx},\, d_+ \hat g_{yy},\,d_+ \hat g_{zz},\, \hat g_{xx},\,  \hat g_{yy},\, \hat g_{zz},\, F_z,\, A\}.
\label{set2}
\end{align}
with corresponding sets of boundary initial data, 
\begin{align}
I_1&=\{ f_x^4,\, f_y^4\},
\label{in1}
\\
I_0&=\{ f^4_z,\, a^4\}.
\label{in2}
\end{align}
The set $S_1$ consists of those bulk fields that vanish at order $\mathcal{O}(\epsilon^0)$, whereas $S_0$ represents all bulk fields which are non-zero at order $\mathcal{O}(\epsilon^0)$. In Appendix \ref{decoupling} we show that $\mathcal{O}(\epsilon)$ terms of  functions in $S_1$ can be computed once we know the $\partial_{\bot}$-derivatives of  the $\mathcal{O}(\epsilon^0)$ contributions to $S_0$ (the planar solutions) and $I_1$. They do not depend on $\mathcal{O}(\epsilon)$ contributions to functions of $S_0$. At the same time  the differential equations for the $\mathcal{O}(\epsilon)$ correction terms of the functions in $S_0$ do not depend on functions in $S_1$ nor on transverse derivatives of functions in $S_0$. \\ \indent 
Remarkably, the first order transverse derivative corrections to the zeroth order solutions of the fields in $S_0$ contribute negligibly  to the expectation value of boundary observables. Their contributions are two orders of magnitude smaller than those of first order corrections to fields contained in set $S_1$. As shown in Appendix \ref{decoupling}, through first order in transverse derivatives, the differential equations for the fields in $S_0$ and the functions in $I_0$   do not contain any explicit transverse derivatives. The only first order corrections to these differential equations  are a consequence of negligibly small $\mathcal{O}(\epsilon)$-corrections to the initial conditions $\{(\hat{g}_{ij})^{(1)}, (a^4)^{(1)}, (f^4_z)^{(1)} \}$ on the first time slice, a result of small transverse gradients appearing in the coordinate transformation from FG to EF coordinates. Thus, up to an error of about $1\%$ we can ignore their first order corrections to the boundary stress energy tensor. Since their equations also decouple from the differential equations for fields in $S_1$ and temporal differential equations of functions in  $I_1$, we can substantially simplify the calculation. In order to compute the relevant contributions to the boundary stress energy tensor up to the first order in transverse derivatives, it is thus sufficient to solve planar shockwave collisions plus only four additional differential equations on each time slice. Those four equations are given by the $\mathcal{O}(\epsilon)$-terms of the $x,y$ components of Eq. (\ref{E:FEOM}), and the $\{x,z\}$ and $\{y,z\}$ components of Eq. (\ref{E:d0gEOM})  given in Appendix \ref{Einstein_equations}. For each point in the transverse plain the numerics are, therefore, only slightly
slower than planar shockwave collisions\iffalse, while the results shown in the coming section \ref{res} agree  well with the exact solutions which can be found in \cite{Che}\fi. It should be added that the approximation of expanding in transverse gradients improves as one further decreases the longitudinal thickness of the shocks or more generally increases the aspect ratio. The parameters in \cite{Che} were chosen to simulate the collision of rather broad blobs of energy; for collisions of more realistic, highly Lorentz-contracted projectiles, the agreement should, thus, further improve. \\ \indent We have computed collisions with identical initial conditions comparing the full first order in transverse derivatives with the above simplified approach. We verified that the contributions to boundary observables in both treatments were identical up to errors of $\approx 1 \%$.
\subsection{Easy  $\mathcal{O}(\nabla_{\bot}^2)$-terms}
\label{partial}  At late times, well after the hydrodynamization time, differences between the exact solution and the approximation by transverse derivative expansion, truncated at first order, become substantial (see e.g. Fig \ref{pressure}, \ref{pressure2}, \ref{energy_3}). At  time $t \approx 4$ higher order corrections in transverse derivatives  can no longer be neglected. However, treating  second order corrections exactly would more than double memory usage and substantially increase run time to a point where the advantage over the exact solution might  become small. Solving radial differential equations and the elliptic differential equation, to update the shift parameter, continue to be much faster compared with the exact treatment. Computing the source terms of each radial differential equation will become the new run-time bottleneck when computing higher order corrections. \\ \indent One can separate second order corrections in transverse derivatives to the boundary stress energy tensor into two categories: contributions from bulk fields and contributions from boundary data which is time evolved using the covariant conservation equation of the stress energy tensor. As explained above, calculating contributions to boundary observables coming from spatial components of the bulk metric is costly in memory  and run time. On the other hand, one may easily determine the partial second order corrections in transverse derivatives to the energy density $(\Delta a^4)^{(2)}$ and the longitudinal momentum density $(\Delta f^4_z)^{(2)}$ arising from the first order terms  of the transverse energy flux $T^{tx}$, $T^{ty}$, the shear components $T^{zx}$, $T^{zy}$  and the conservation equation of the boundary stress energy tensor (\ref{time_der_energy_(flux)}) using
\begin{align}
\label{boundary_eq_1}
\partial_t \, (\Delta a^4)^{(2)}&= \frac{2}{3}\Big(\partial_x (f^4_x)^{(1)}+\partial_y (f^4_y)^{(1)}+\partial_z (\Delta f^4_z)^{(2)}\Big), \\ \partial_t  (\Delta f^4_z)^{(2)} & = \frac{1}{2}\partial_z (\Delta a^4)^{(2)}-\partial_x (\hat g_{xz}^4)^{(1)}-\partial_y (\hat g_{yz}^4)^{(1)} ,
\label{boundary_eq_2}
\end{align}
plus the initial condition $(\Delta f^4_z)^{(2)}= (\Delta a^4)^{(2)}=0$ on the first time slice. Solving for this subset of second order corrections requires negligible additional memory and wall-clock time. \iffalse However, unlike at first order in transverse derivatives, where we could split up the set of bulk fields into two sets with relevant and irrelevant contributions to boundary expectation values of the stress energy tensor and ignore the negligible ones, these second order corrections are not necessarily the dominant contributions at this order.\fi \\ \indent A priori, one might only expect to obtain a qualitative estimate of the size of second order corrections from such an incomplete calculation of second order terms. Hence, it is all the more remarkable that we find that inclusion of second order terms in the boundary conservation equation (\ref{boundary_eq_1}) and (\ref{boundary_eq_2}), alone, drastically improves the agreement between the transverse derivative expansion and the exact result for most observables (see Fig. \ref{pressure2}, \ref{energy_2}, \ref{energy_3}).  Nonetheless, this partial second order treatment should be employed with care, like every partial higher order calculation without proof one has captured a dominating contribution. \iffalse  \\ \indent The $\mathcal{O}(\epsilon^2)$-corrections to initial conditions and the spatial components of the bulk metric are comparatively  expensive to compute regarding wall-clock time. However, from the $\mathcal{O}(\epsilon)$-terms of the transverse energy flux $T^{tx}$, $T^{ty}$ and the conservation equation of the boundary stress energy tensor (\ref{time_der_energy_(flux)})  we can easily determine the following $\mathcal{O}(\epsilon^2)$-corrections  to the energy density $(\Delta a^4)^{(2)}$ and the longitudinal momentum density $(\Delta f^4_z)^{(2)}$
\begin{equation}
\partial_t \, (\Delta a^4)^{(2)}= \frac{2}{3}\Big(\partial_x (f^4_x)^{(1)}+\partial_y (f^4_y)^{(1)}+\partial_z (\Delta f^4_z)^{(2)}\Big), \,\,\,\,\, \partial_t  (\Delta f^4_z)^{(2)} = \frac{1}{2}\partial_z (\Delta a^4)^{(2)},
\end{equation}
where we have used the result of the previous section $( a^4)^{(1)}\approx 0$.\fi
\section{Results}
\label{res}
\begin{figure}
\begin{center}
\includegraphics[scale=0.75]{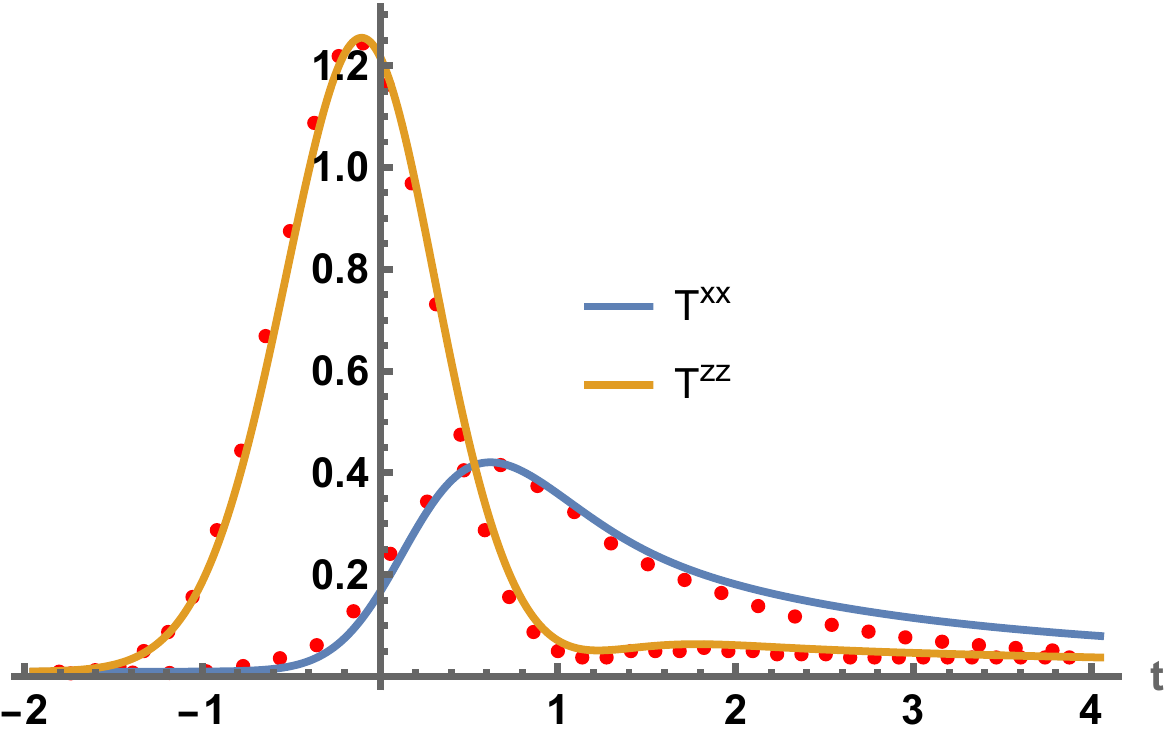}
\end{center}
\caption{The $x$ and $z$ components of the pressure,  $T^{zz}$ and $T^{xx}$, at  $\mathcal{O}(\epsilon)$, shown as a function of time at $x=y=z=0$. Again we see good agreement with the results in  \cite{Che} (red dots),  slowly deteriorating for  $T^{xx}$ at late times.}
\label{pressure}
\end{figure}

\begin{figure}
\begin{center}
\includegraphics[scale=0.65]{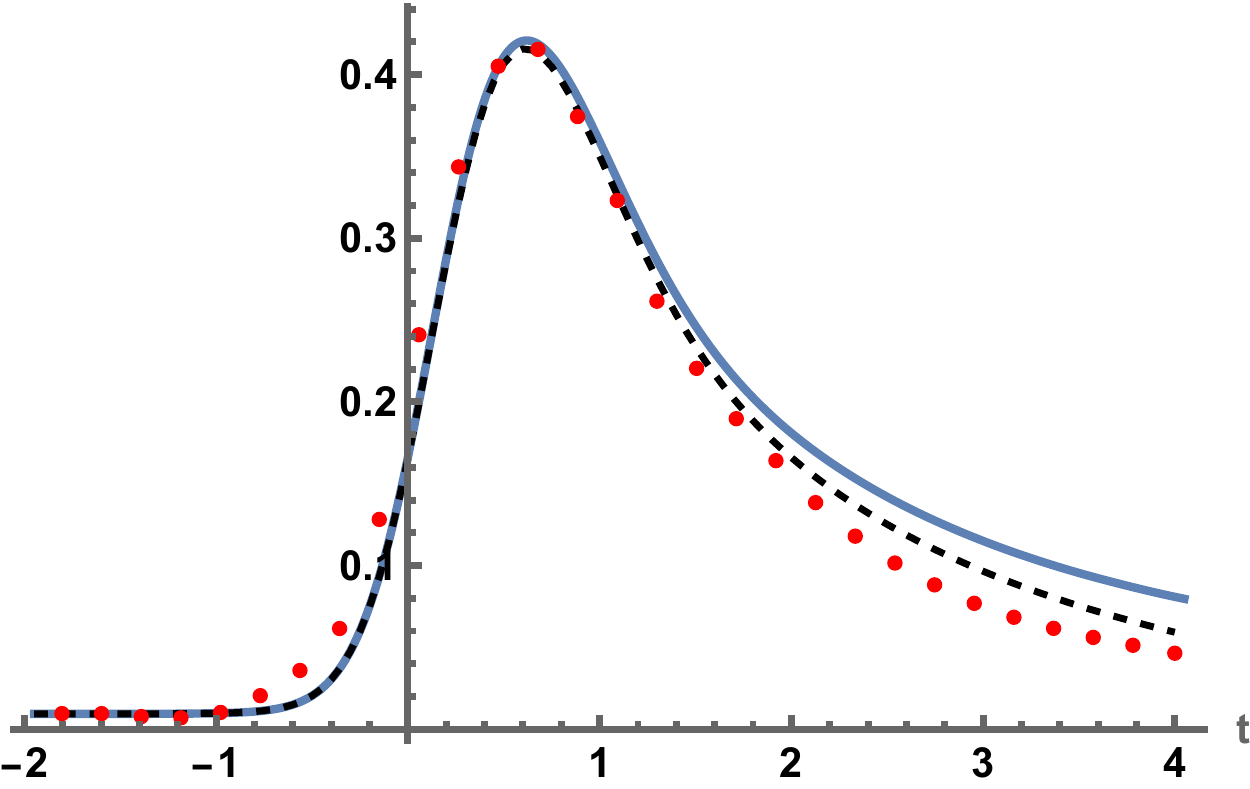}
\end{center}
\caption{The $x$-component  of the pressure   $T^{xx}$ up to order  $\mathcal{O}(\epsilon)$ (solid blue line), shown as a function of time at $x=y=z=0$. Again the red dots represent the exact result from  \cite{Che}. The black dashed line includes partial $\mathcal{O}(\epsilon^2)$-corrections discussed in section \ref{partial}.}
\label{pressure2}
\end{figure}

\begin{figure}
\begin{center}
\includegraphics[scale=0.57]{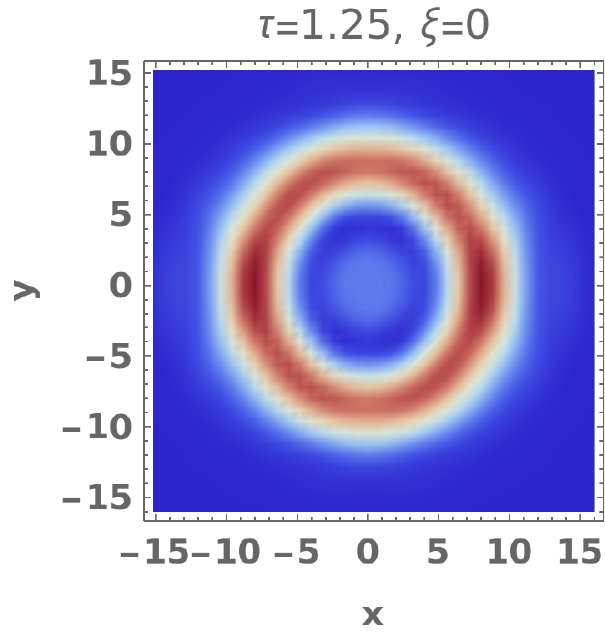}
\includegraphics[scale=0.575]{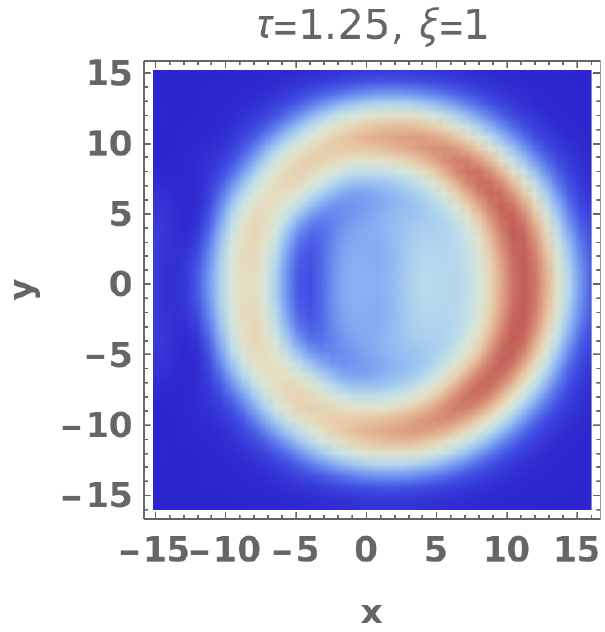}
\includegraphics[scale=0.555]{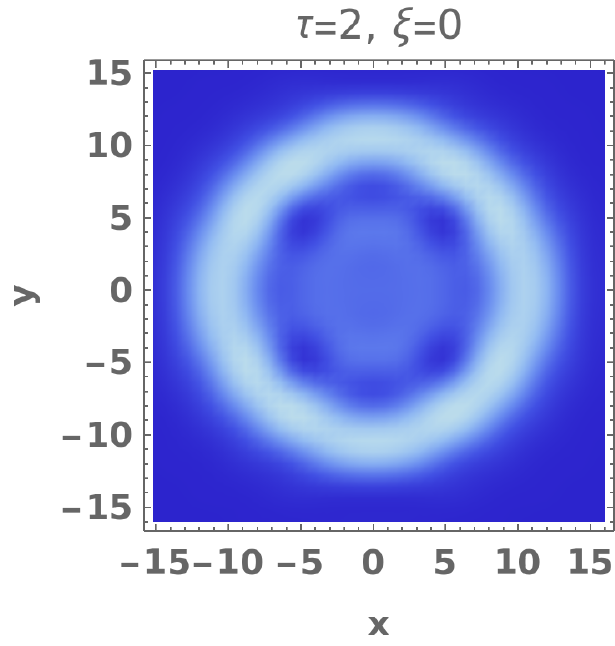}
\includegraphics[scale=0.55]{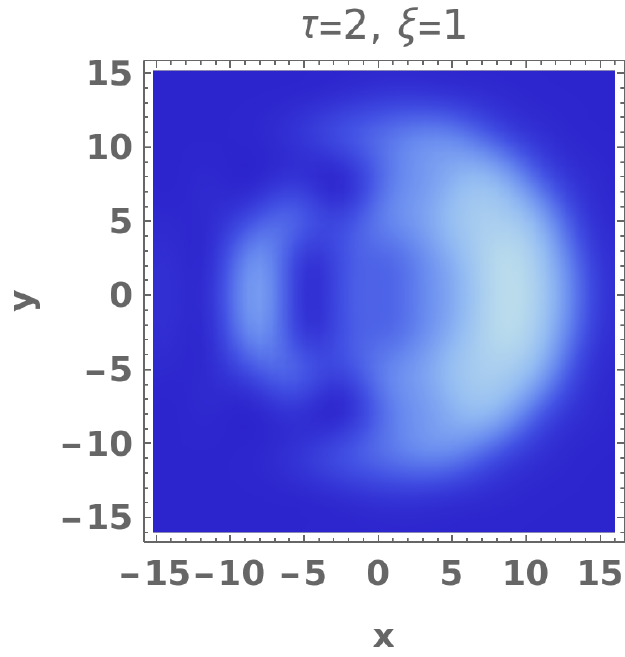}
\includegraphics[scale=0.45]{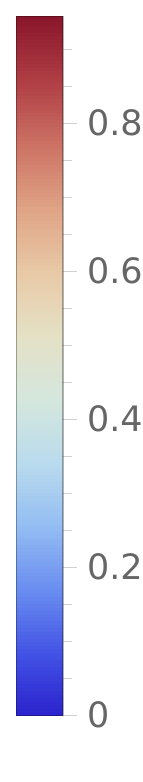}
\end{center}
\caption{The hydrodynamic  residual $\Delta$ at zeroth order in transverse derivatives, for proper times $\tau = 1.25$, $\tau = 2$ and rapidities $\xi=0$, $\xi=1$. Already at zeroth order the residual $\Delta^{(0)}$ agrees well with the exact results. As in \cite{Che} we find a hydrodynamization time of about $t \approx 1.25$ for the central region. }
\label{resudial_fig_0}
\end{figure}

\begin{figure}
\begin{center}
\includegraphics[scale=0.525]{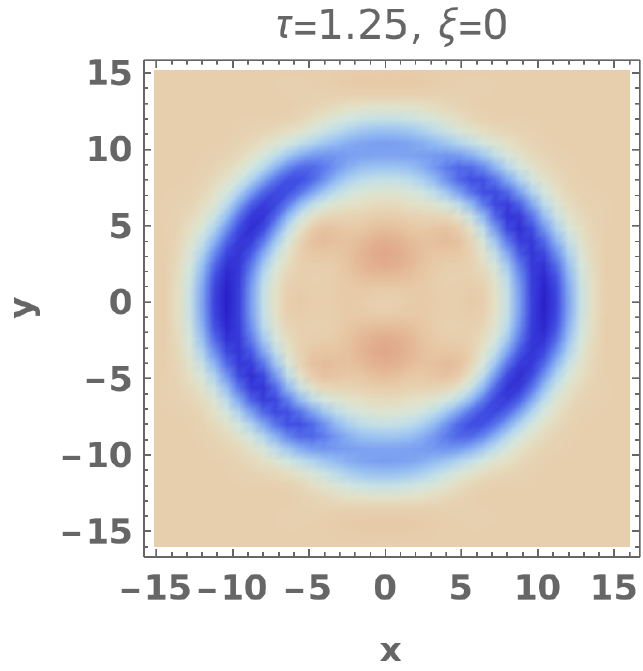}
\includegraphics[scale=0.525]{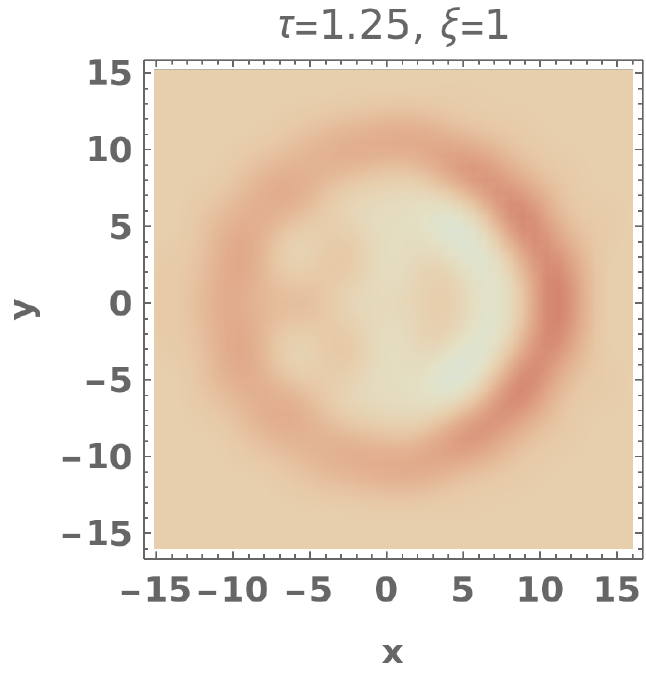}
\includegraphics[scale=0.525]{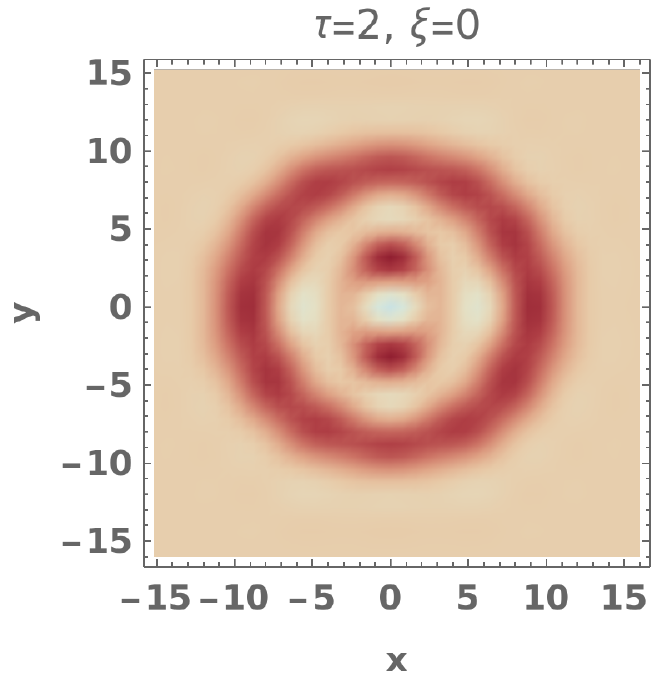}
\includegraphics[scale=0.525]{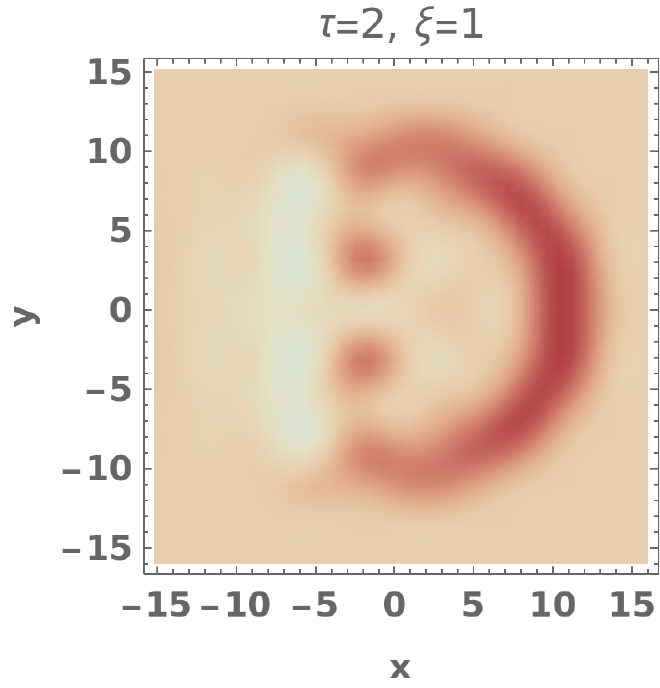}
\includegraphics[scale=0.45]{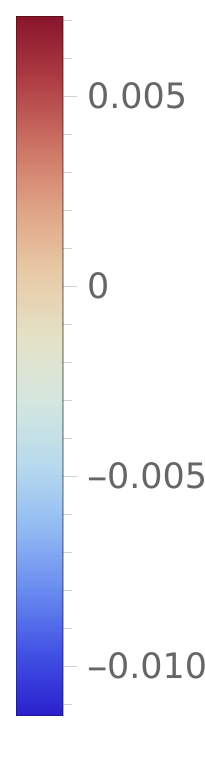}
\end{center}
\caption{First order corrections in transverse derivatives to the hydrodynamic residual $\Delta$ at proper times $\tau = 1.25$, $\tau = 2$ and rapidities $\xi=0$, $\xi=1$. Consistent with the agreement between the results at zeroth order and the exact results in \cite{Che}, we find  the first order correction to the residual $\Delta$ to be less than $2\%$. This is  a consequence of the fact that only fields in $S_0$  given in  (\ref{set2})  contribute to the residual at first order.}
\label{resudial_fig_1}
\end{figure}

\begin{figure}
\begin{center}
\includegraphics[scale=0.9]{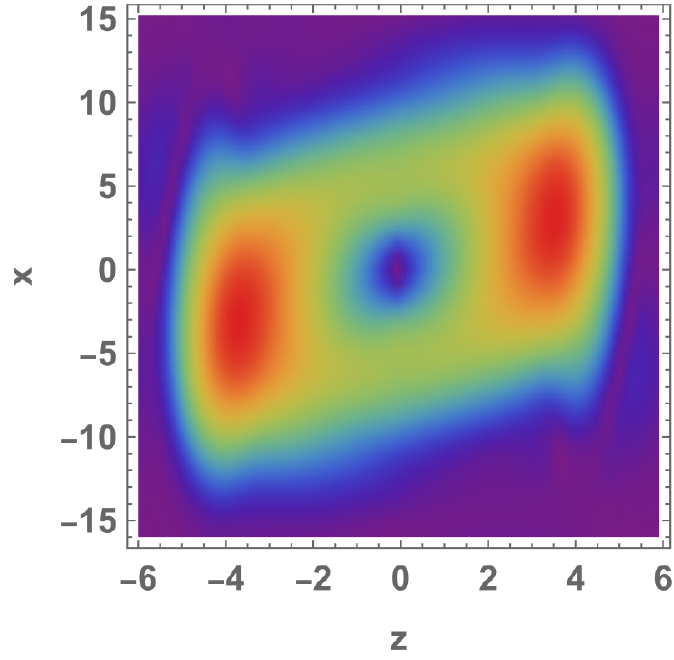}
\includegraphics[scale=0.9]{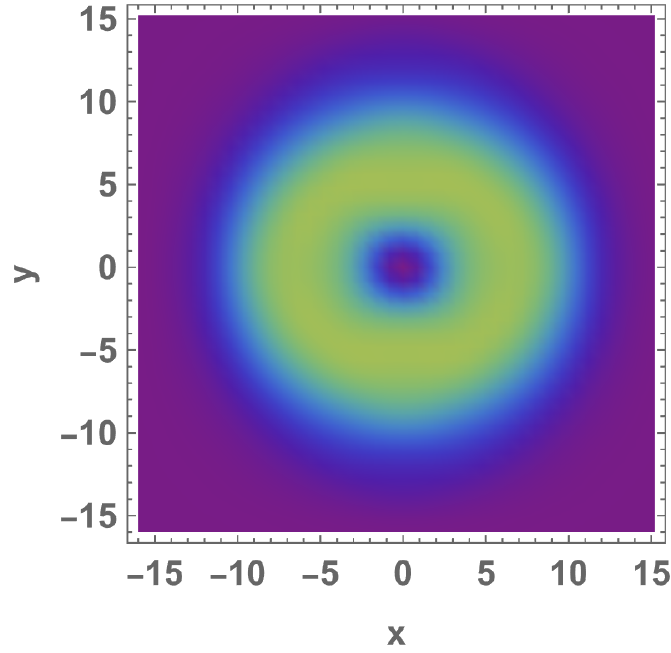}
\includegraphics[scale=0.65]{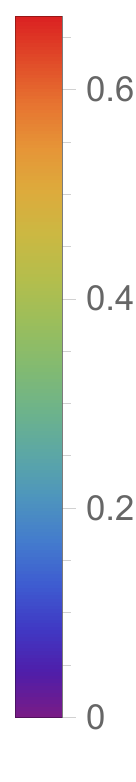}
\end{center}
\caption{The fluid velocity $|\bold v|=| \bold u/u^{0}|$ at $t=4$ up to first order in transverse derivatives. On the left hand side we display the $y=0$ slice of $|\bold v|$,  while the right side shows the $z=0$ slice.}
\label{fluid_velocity}
\end{figure}

\begin{figure}
\begin{center}
\includegraphics[scale=0.95]{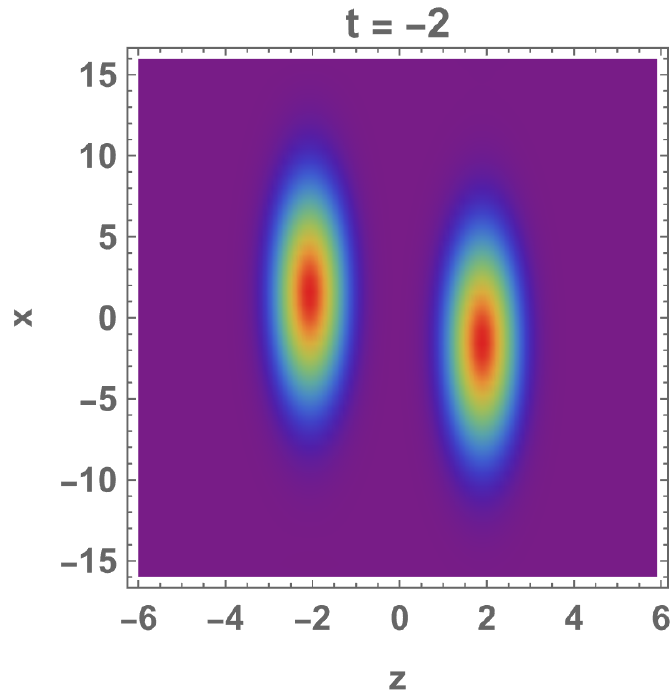}
\includegraphics[scale=0.65]{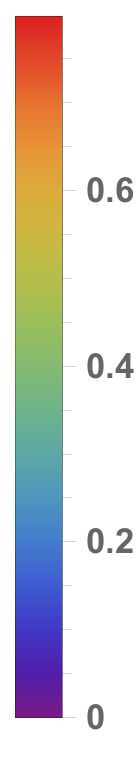}
\includegraphics[scale=0.95]{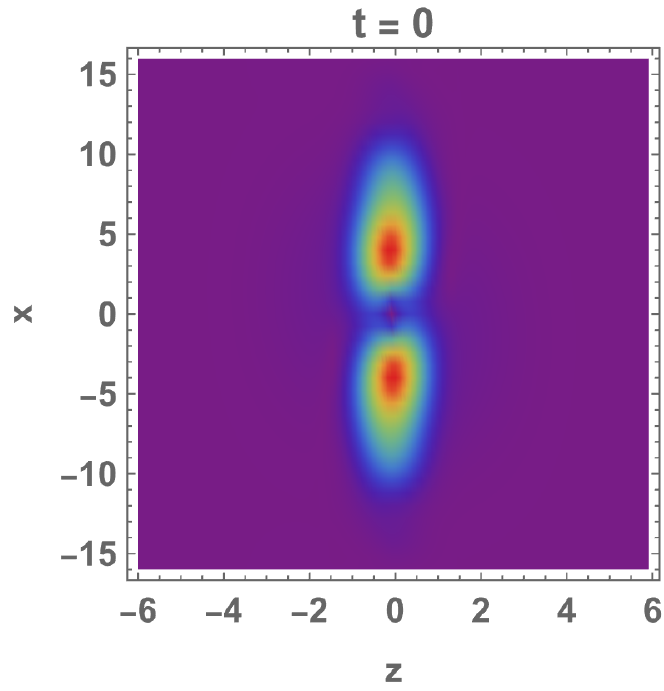}
\includegraphics[scale=0.65]{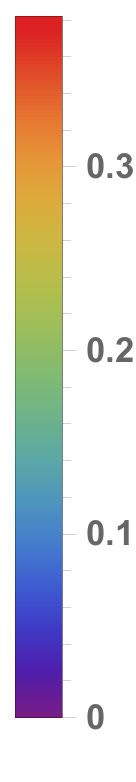}

\includegraphics[scale=0.95]{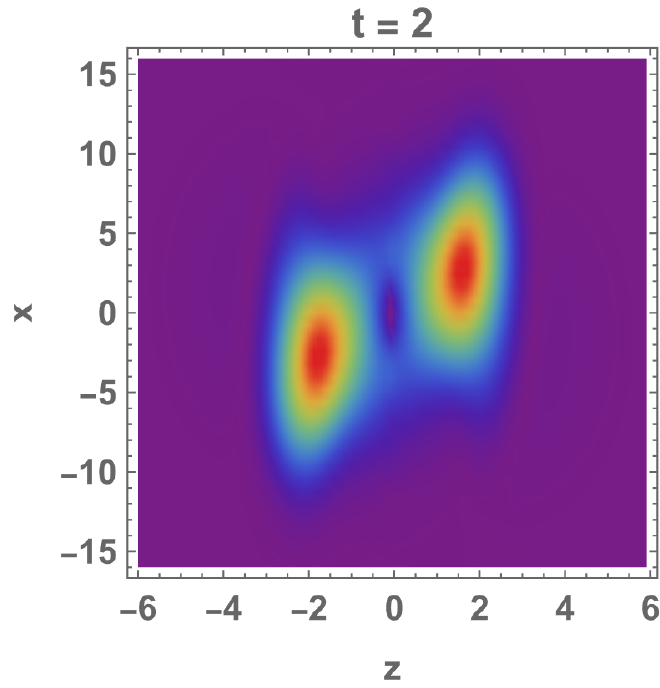}
\includegraphics[scale=0.65]{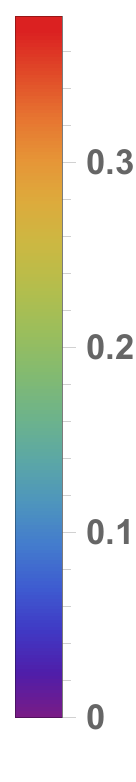}
\includegraphics[scale=0.95]{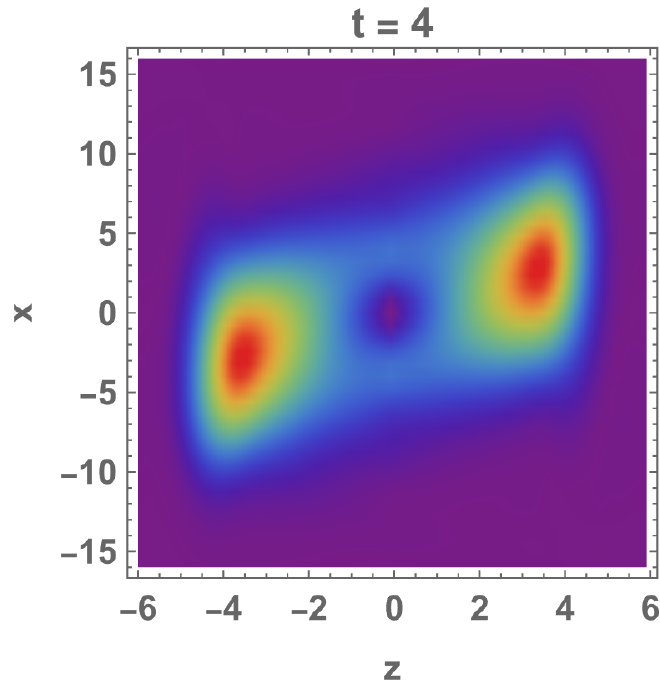}
\includegraphics[scale=0.65]{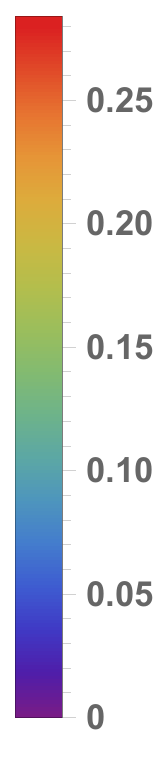}
\end{center}
\caption{The energy flux $|T^{0i}|$ including first order corrections in transverse derivatives, for times  $t=-2$, $t=0$, $t=2$ and $t=4$ (from top left to bottom right), and at $y=0$.}
\label{momentum}
\end{figure}

\begin{figure}
\begin{center}
\includegraphics[scale=0.95]{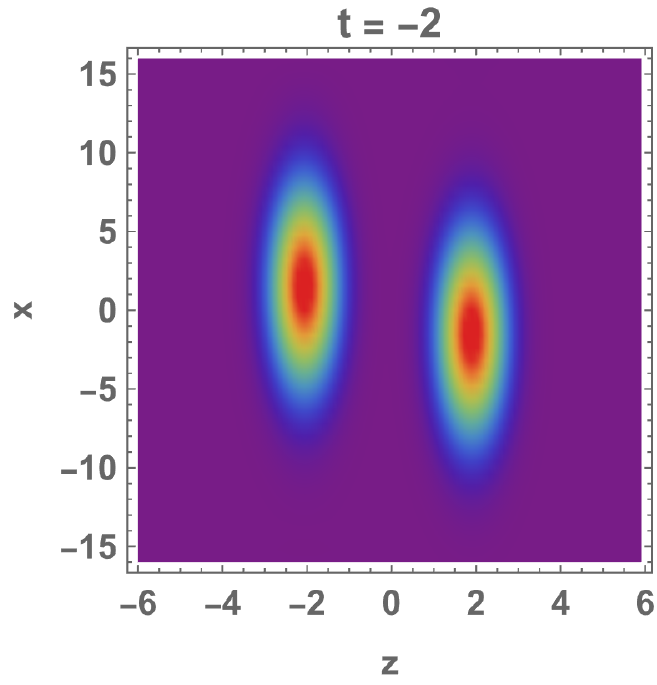}
\includegraphics[scale=0.65]{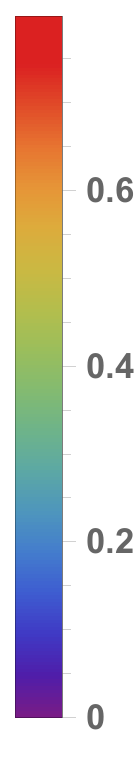}
\includegraphics[scale=0.95]{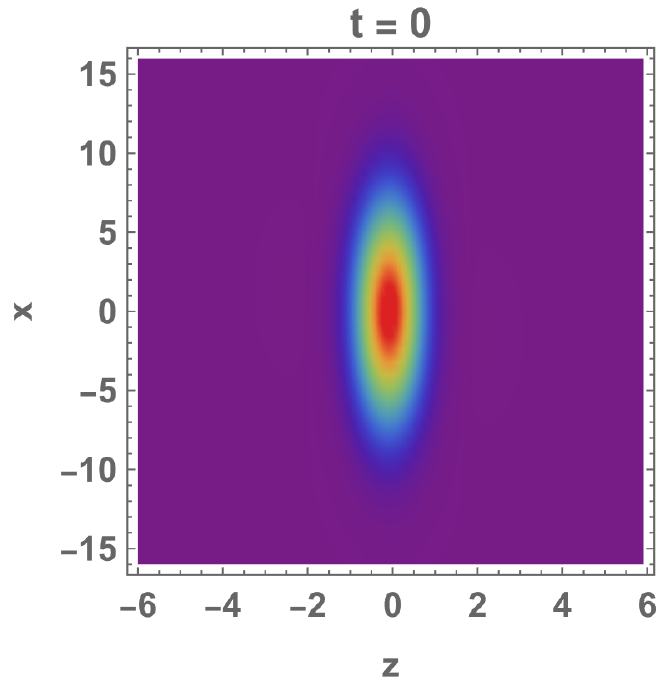}
\includegraphics[scale=0.65]{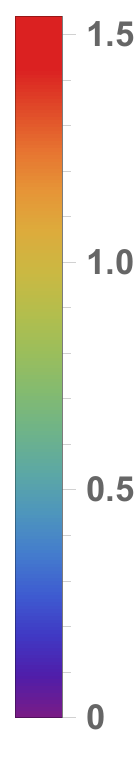}

\includegraphics[scale=0.95]{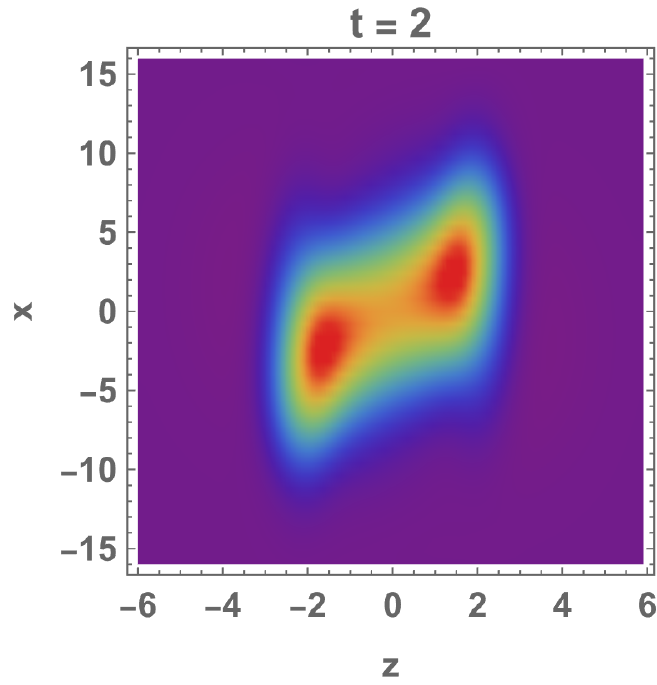}
\includegraphics[scale=0.65]{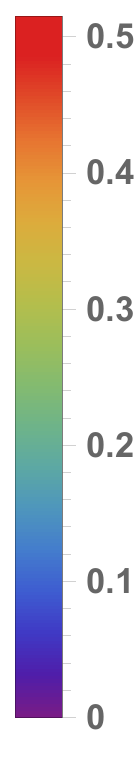}
\includegraphics[scale=0.95]{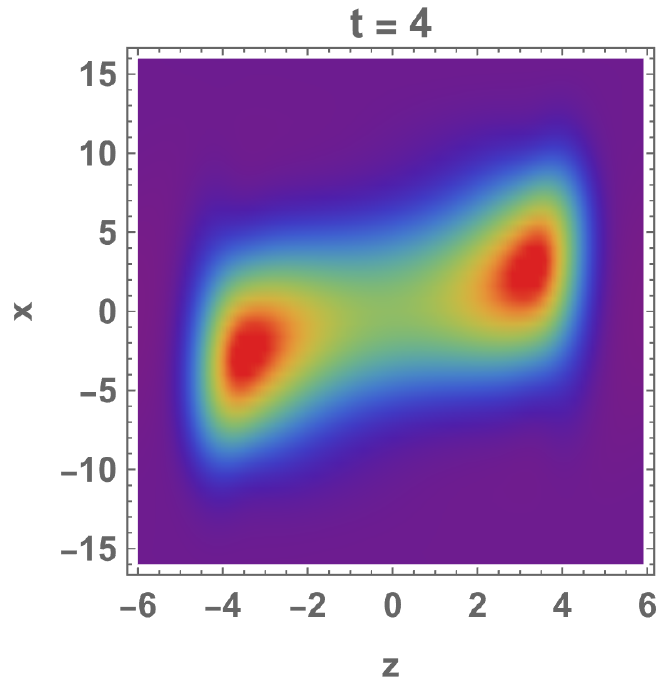}
\includegraphics[scale=0.65]{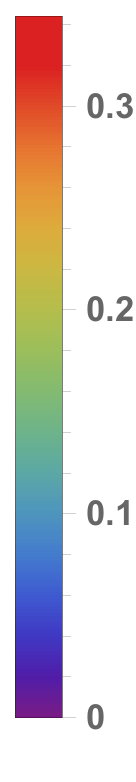}
\end{center}
\caption{The energy density $T^{00}$ at zeroth order,  for times  $t=-2$, $t=0$, $t=2$ and $t=4$ (from top left to bottom right), and at $y=0$.  First order corrections to $T^{00}$ are negligible.}
\label{energy_1}
\end{figure}

\begin{figure}
\begin{center}
\includegraphics[scale=0.57]{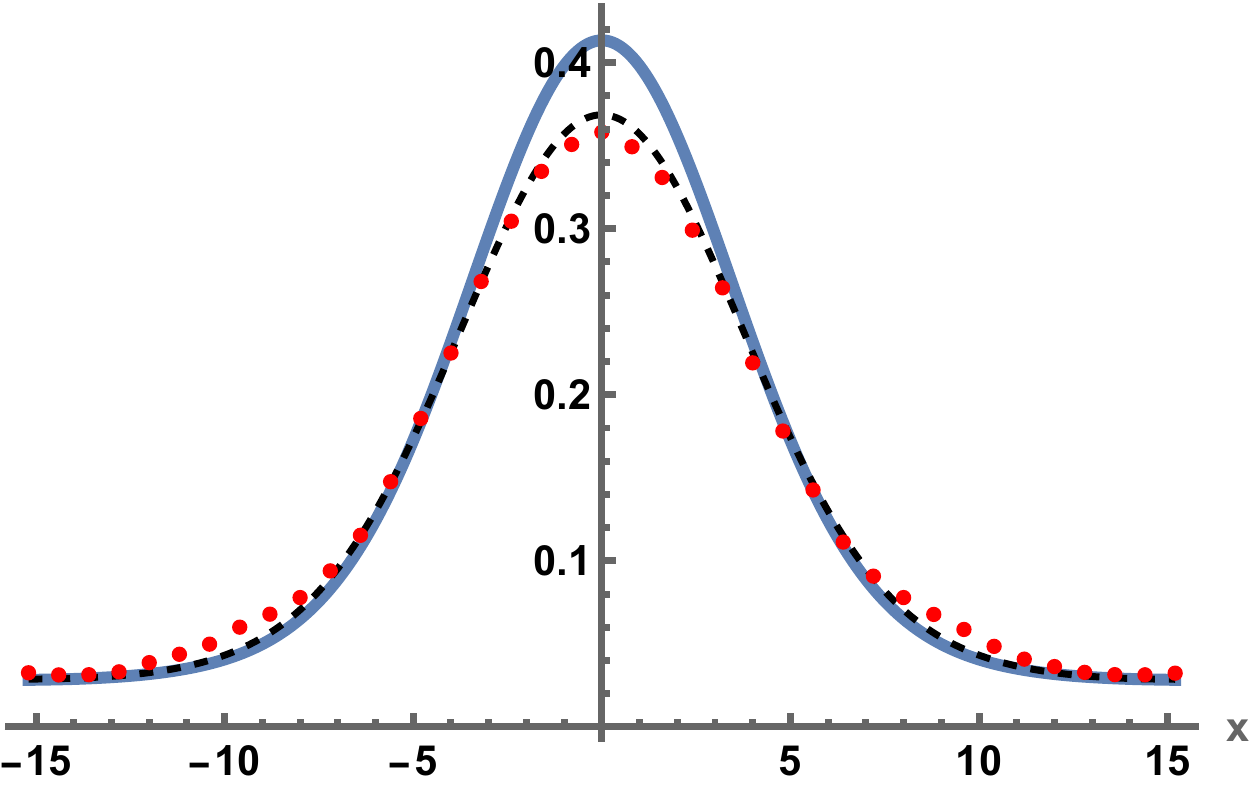}
\includegraphics[scale=0.57]{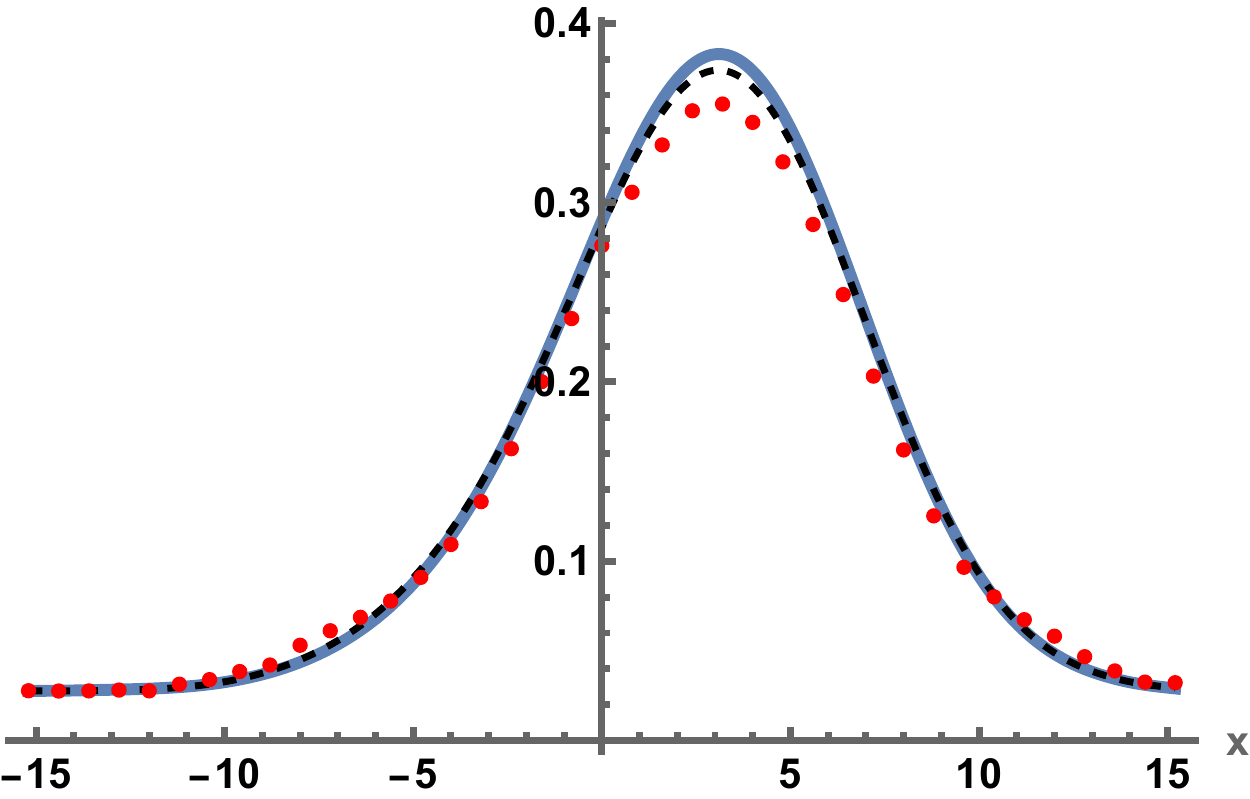}
\end{center}
\caption{The energy density at zeroth order  (solid blue curve)  in the central region $y=0$, $z=0$ (left) as well as at $z=2.1$ (right) at time $t=2$. The red dotted curve represents the exact results, while the black dashed curve includes the partial $\mathcal{O}(\epsilon^2)$ corrections discussed in section \ref{partial}.}
\label{energy_2}
\end{figure}

\begin{figure}
\begin{center}
\includegraphics[scale=0.57]{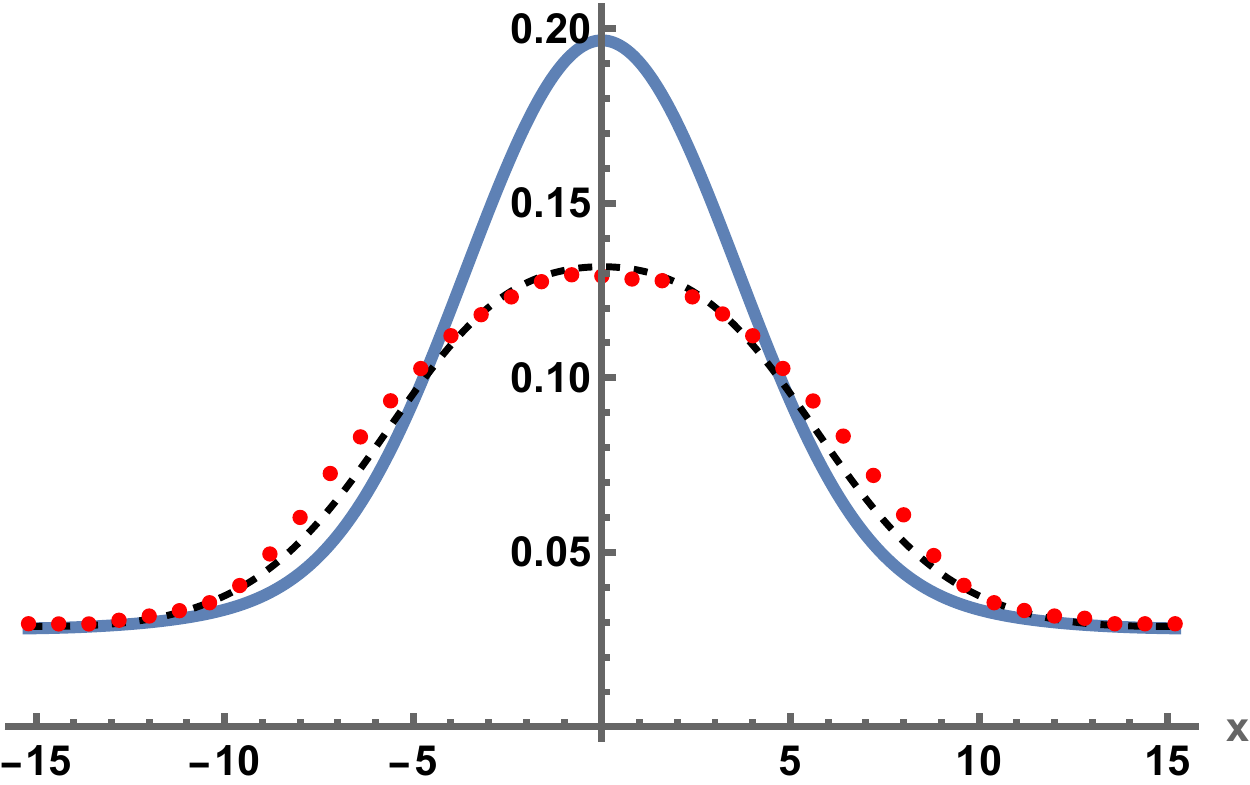}
\includegraphics[scale=0.57]{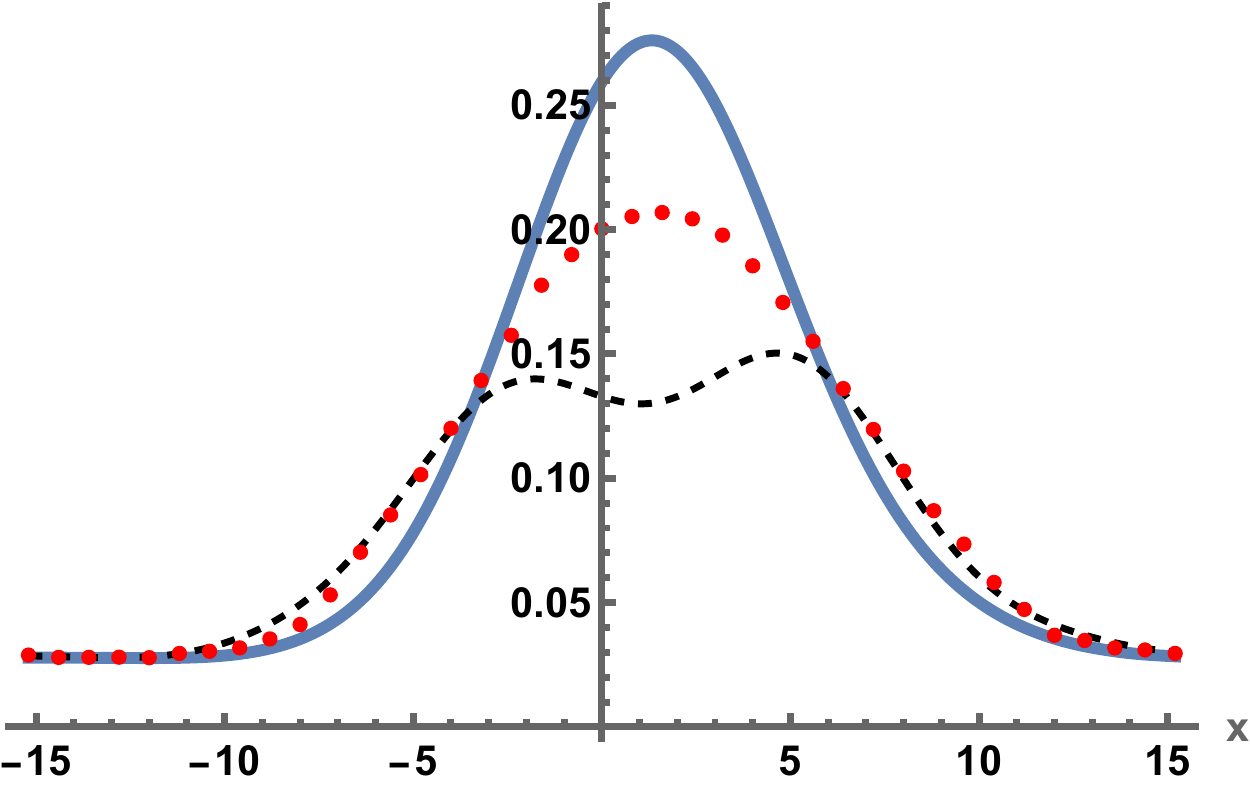}
\includegraphics[scale=0.57]{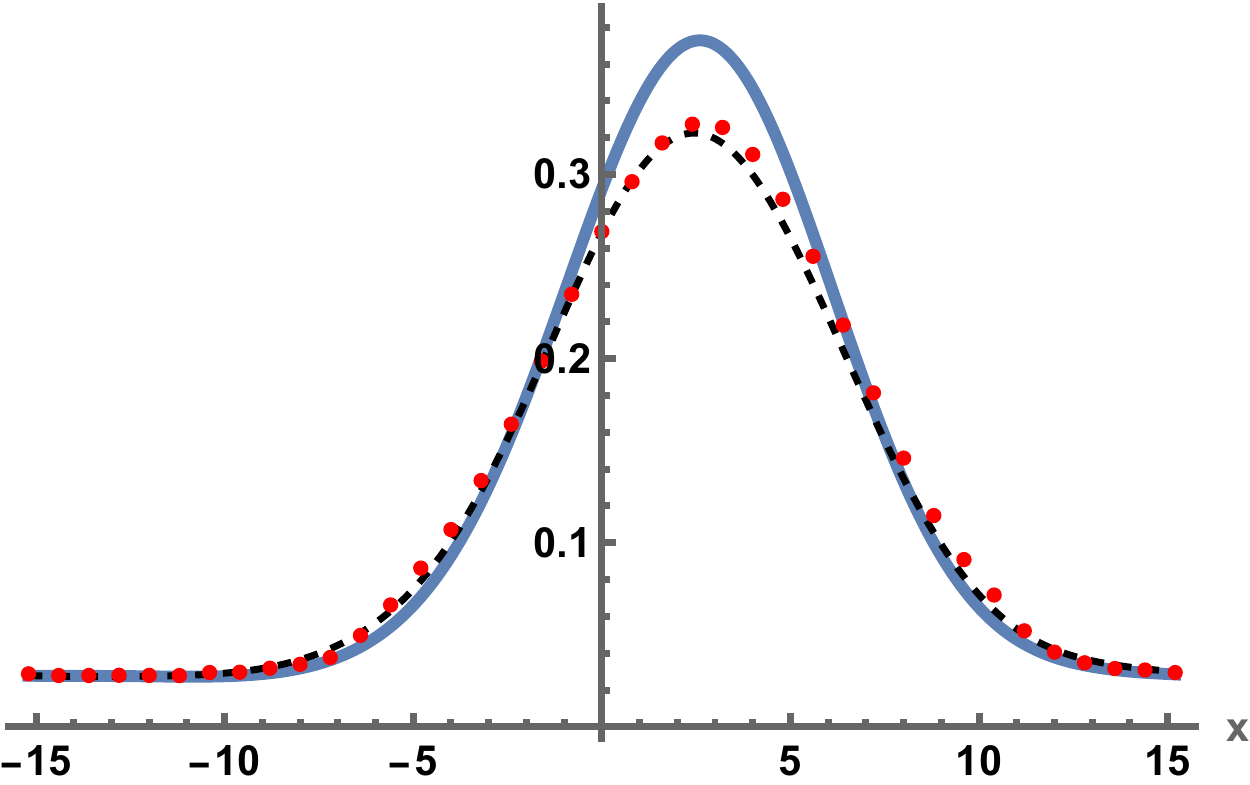}
\end{center}
\caption{The energy density at zeroth order (solid blue curve)  in the central region $y=0$, $z=0$ (top left) and   at $z=2.1$ (top right) and  $z=3.3$ (bottom)  at time $t=4$. The red dotted curve represents the exact results, while the black dashed curve includes partial $\mathcal{O}(\epsilon^2)$ corrections discussed in section \ref{partial}. Differences between the exact solution and the truncated transverse derivative expansion continue to increase with time, such that long after the hydrodynamization time they become substantial. Therefore, it appears to be a fluke that the (partial) second order corrected solution and the exact solution for vanishing $z$ fit so well in the top left plot.}
\label{energy_3}
\end{figure}

We compute the coordinate transformation from FG to EF coordinates using a Chebyshev grid  in the radial direction with three domains containing $31$ grid points each and Fourier grids in longitudinal and transverse directions. The longitudinal grid consists of $N=256$ and the transverse grids consist of $N_{x,y}=36$ grid points. For the time evolution in EF coordinates we found a radial Chebyshev grid with only two domains with $18$ grid points each to be sufficient. After computing the EF initial data, to perform the time evolution we remap data to a smaller transverse Fourier grid with $12$ grid points in each dimension and a coarser longitudinal  Fourier grid with $140$ grid points. To enable comparison of our approximation with the exact results our physical parameters are chosen to match those  in \cite{Che}. The physical sizes of the spatial $3$-torus in which we placed the shocks are $L_z=12$ in longitudinal direction and $L_{x,y}=32$ in transverse direction. We chose the longitudinal width parameters of the shocks to be  $w=0.5$ and $R=4$ and chose a unit amplitude  $\mathcal{A}=1$. The impact parameter is $\bold b = \frac{3}{4} R \, \hat{\bold x}$.  We perform our calculation with a background energy density of $3.6\%$ of the energy density's  peak value. Discretizing non-linear PDEs introduces artificial, high frequency oscillations that have to be removed via numerical filters. The details of this filtering procedure is discussed in \cite{che3}. On each time slice we apply a filter in longitudinal direction on the updates  of the spatial components of the bulk metric, the energy density $a^4$, the momentum density $f_i^4$ and the shift parameter $\lambda$ (all of which we compute via the fourth order Runge-Kutta method), removing one third of all modes corresponding to the highest frequencies, while keeping the remaining two thirds. In addition we apply a radial filter on the longitudinally filtered updates by  transforming the radial Chebyshev grid onto a slightly smaller, single domain grid with $35$ grid points, and transforming back to the original grid. We do not filter during the individual substeps of the Runge-Kutta algorithm.
\\ \indent We compare the stress energy tensor after the collision with its hydrodynamic approximation, where the constitutive relations are truncated after the first order in derivatives. For this we solve the Eigenvalue equation
\begin{equation}
T^\mu\,_\nu \,u^\nu = -\varepsilon u^\mu
\end{equation}
order by order in transverse derivatives for the fluid velocity $u^\mu$ and the proper energy density $\varepsilon$. \iffalse Note that we use the same symbol $\epsilon$ for the proper energy density and the expansion parameter $\epsilon$ whose powers count powers of transverse derivatives. Henceforth they can be distinguished by the context in which they are used. \fi
The hydrodynamic approximation 
\begin{equation}
    \widehat T^{\mu\nu}_{\rm hydro}
    = p \, g^{\mu\nu} + (\varepsilon{+}p) \, u^\mu u^\nu + \Pi^{\mu \nu} \,,
    \label{hydro_approx}
\end{equation}
where the viscous stress is given by
\begin{equation}
    \Pi_{\mu\nu}
    =
    -2\,\eta \,
    \left[
    \partial_{(\mu}u_{\nu)}
    + u_{(\mu} u^\rho \partial_\rho u_{\nu)}
    - \tfrac 13 \,  \partial_\alpha u^\alpha
    (\eta_{\mu\nu} + u_\mu u_\nu)
    \right]
    + \mathcal{O}(\partial^2) \,,
\end{equation}
is also expanded up to order $\mathcal{O}(\nabla_{\bot})$. Here $p$ is the pressure \iffalse\footnote{ Note that $(T_{xx})^{(i)}=(T_{yy})^{(i)}$ for $i=0,1$.}\fi and $\eta$ the shear viscosity. In order to evaluate how well the system is described by hydrodynamics we compute the residual
\begin{equation}
\Delta = \frac{3}{\varepsilon} \sqrt{\Delta T^{\mu \nu} \Delta T_{\mu \nu}}
\label{residual}
\end{equation}
with $\Delta T^{\mu \nu} = T^{\mu \nu}- \widehat T^{\mu\nu}_{\rm hydro}$. Again let $\Delta^{(i)}$ denote the  $\mathcal{O}(\epsilon^i)$ contribution to $\Delta$. The hydrodynamic approximation is generally taken to be valid for $\Delta< 0.15$. \iffalse  For planar shockwave collisions, both symmetric and asymmetric, it is known that this threshold is reached roughly at proper time $\tau\approx 2$ as shown in \cite{che3, Chesler:2015fpa,wae3}.\fi We show the results for the residual at zeroth order, $\Delta^{(0)}$,  and the first order  correction $\Delta^{(1)}$, at proper times  $\tau =1.25$, $\tau =2$  and rapidities $\xi=0$, $\xi = 1$ in Fig. \ref{resudial_fig_0} and  Fig. \ref{resudial_fig_1}, respectively. Evidently  the inclusion of first order transverse dynamics only has a small effect on the residual $\Delta$,  and  already at zeroth order in transverse derivatives we find  in the central region (small $\bold x_\bot$) the same  hydrodynamization time\footnote{\iffalse
The difference between the hydrodynamization time of the central region at vanishing transverse derivatives and the hydrdodynamization time found in planar shock collisions \cite{che3,Chesler:2015fpa,wae3} is in part explained by a  different amplitude. Another cause is a small 'hydrodynamization island', presumably an artefact of insufficient longitudinal resolution. The island with $\Delta < 0.15$ starts at about $t \approx 1.25$ and extends to the low rapidity region of the physical relevant hydrodynamization proper time (approximate) hyperbola $\sqrt{(t-0.58)^2-z^2}=1.5$. For times  $2 \lesssim t$ we find the connected region with $\Delta<0.15$ to be again bounded by a hyperbola of constant proper time.\fi This result is highly sensitive to the choice of the threshold $\Delta < 0.15$ at which we consider the hydrodynamic description to be valid, with a potential residual sensitivity to the longitudinal grid size.} of $t\approx 1.25$ as in \cite{Che}.  \\ \indent  Our results regarding transverse energy flux, shown earlier  in  Fig. \ref{transv} and \ref{transvmax}, illustrate  good agreement  between the  $\mathcal{O}(\epsilon)$-approximation and the exact solution regarding transverse flow.  In Fig. \ref{pressure} we display the $xx$ and $zz$-components of the stress  tensor, $T^{xx}$ and $T^{zz}$, at the origin $x=y=z=0$ as a function of time. Again both stress components agree well with the results in \cite{Che}. Only for $T^{xx}$ and at late times, $t>2$, do we observe a noticeable disagreement. As shown in Fig. \ref{pressure2}, this error is more than halved by including the partial  $\mathcal{O}(\epsilon^2)$-corrections discussed in section \ref{partial}. This suggests that even for rather broad shocks, as shown in this work, and late times $t \approx 4$ the full inclusion of $\mathcal{O}(\epsilon^2)$-corrections should provide a decent approximation, whereas for earlier times, $t<2$, first order corrections appear sufficient.\\ \indent  We computed the  fluid 3-velocity $\bold v = \bar u/u^0$ up to order $\mathcal{O}(\epsilon)$ and display its absolute value at time $t=4$ on both a $x=0$ slice and a $z=0$ slice in Fig. \ref{fluid_velocity}. In Fig. \ref{momentum} we display the absolute energy flux $|T^{0i}|$ at various times on the $y=0$ slice. In Fig. \ref{energy_1} we show the analogous plots for the energy density $T^{00}$. A general comparison of $\mathcal{O}(\epsilon)$-results and exact results shows that the worst agreement occurs  for the energy density $T^{00}$ at late times $t=4$. Therefore we study in Fig. \ref{energy_2} and  \ref{energy_3} the energy density on several $z=const$, $y=0$ slices at times $t=2$ and $t=4$ respectively. At time $t=2$ we still observe quite good agreement with the exact results, with  partial  $\mathcal{O}(\epsilon^2)$-corrections again improving the approximation. At  $t=4$ the errors appear to be substantial. However, including partial  $\mathcal{O}(\epsilon^2)$-corrections  again substantially improves agreement in the central region $z=0$ and for large $z$. However, at intermediate values for the longitudinal coordinate $z \approx 2$ a notable mismatch between approximation and exact results remains.

\iffalse  A possible explanation for why $\mathcal{O}(\nabla^1_{\bold{x}_\bot})$ corrections are not sufficient to satisfyingly reproduce  hydrodynamization times comes from the unequal speed with which different components of the stress energy tensor approach their hydro approximation. At $\tau=1$ and vanishing rapidity the biggest contribution to the residual $\Delta$ comes from the pressure components, whereas e.g. the $00$-component in the central region differs from $\widehat T^{00}_{\rm hydro}$ only negligibly. It sounds plausible that the $T_{xx}$, $T_{yy}$ and $T_{xy}$ require us to consider $\mathcal{O}(\nabla^2_{\bold{x}_\bot})$ corrections, in order to be well approximated by an expansion in transverse gradients. 
\fi

\FloatBarrier
\section{Conclusion}
\label{conclusion}
In this work we introduce the framework of a transverse derivative expansion to allow more efficient calculations of holographic models of heavy ion collisions. The first order correction to collisions of planar shocks allows us to reproduce the exact solutions in \cite{Che} remarkably well. The transverse derivative expansion has substantial computational advantages over the very memory and time consuming exact treatment. At a given point in the transverse plane, Green's functions of the differential operators that have to be solved on each time slice are identical to those computed for planar collisions. Transverse derivatives only contribute to source terms. Moreover, it is possible to decouple the Einstein equations at first order in transverse derivatives into two sets. One of them only contributes negligibly to boundary observables. In section \ref{hierarchy} and Appendix section \ref{decoupling} discuss why this happens. As a consequence the equations to be solved on each time slice, and each transverse pixel, up to first order in transverse derivatives reduce to those for planar shock collisions plus only four additional differential equations. This reduces the computation time, compared with the exact treatment of localized shock collisions, roughly by an order of magnitude.\\ \indent In future works we intend to employ the methods described in this paper to study more realistic models of collisions of   heavy ions including a granular structure and transverse energy density fluctuations, which are required  to explain the observed large event-by-event fluctuations of flow observables. As found in  \cite{Che}  there is a considerable transverse flow, which develops before the  hydrodynamic description becomes valid. The question of how sizable transverse energy density fluctuations of the localized shock's initial conditions affect the transverse dynamics before hydrodynamization remains an interesting and open question. \iffalse Especially since it is already known that taking the granular  structure of heavy ions into account does strongly influence the post collision dynamics and hydrodynamization time scale \cite{mue} even for planar shock collisions without transverse dynamics.\fi
\section*{Acknowledgements}
The work of SW was supported by the Feodor Lynen fellowship program of the Alexander von Humboldt foundation. LY and SW acknowledge support by the U.S. Department of Energy grant DE-SC-0011637.
\begin{appendices}
\section{Einstein equations}
\label{Einstein_equations}
We are going to work with the Einstein equations in the chracteristic formulation first derived for AdS space in \cite{che3}. The full form of the equations including the source terms for the constraint equations can be found there in Eq. (3.22)-(3.39). We are going to write explicitly those differential equations which we need to solve on each time slice and leave out constraint equations which are used to update the momentum and energy density:
\begin{subequations}
\label{E:EOM}
\begin{align}
\label{E:SigmaEOM}
	\left(\partial_{r}^2+ \,Q_\Sigma[\hat g] \right) \Sigma &= 0 \\
\label{E:FEOM}
	\left(\delta_i^j \, \partial_{r}^2 +P_{F}[\hat{g},\,\Sigma]_i^j  \, \partial_r +Q_F[\hat{g},\,\Sigma]_i^j\right) F_j &= S_{F}[\hat{g},\,\Sigma]_i \\
\label{E:d0SigmaEOM}
	\left(\partial_{r} + Q_{d_+\Sigma}\right)d_+\Sigma & = S_{d_+\Sigma}[\hat{g},\,\Sigma,\,F]\\
\label{E:d0gEOM}
	\left( \delta^k_{(i}\delta^l_{j)}\partial_r +  Q_{d_+\hat g}[\hat g, \Sigma]^{kl}_{ij}\right)d_+\hat{g}_{kl} & = S_{d_+\hat{g}\,}[\hat{g},\,\Sigma,\,F,\,d_+\Sigma]_{ij}\\
\label{E:AEOM}
	\partial_{r}^2 A &= S_{A}[\hat{g},\,\Sigma,\, F,\,d_+\Sigma,\,d_+\hat{g}].
\end{align}
\end{subequations}
In the following we are going to denote the
connection of the spatial, covariant derivative, modified to be covariant under radial shifts, as $\tilde{\Gamma}^i_{jk}$ and define 
\begin{equation}
\tilde{\Gamma}^i_{jk} \equiv \Gamma^i_{jk}+\frac{G^{il}}{2}(F_k\, G'_{lj}+F_j\, G'_{lk}-F_l\, G'_{jk})
\end{equation}
with the standard Christoffel symbol for covariant derivatives acting on spatial tensor fields
\begin{equation}
\Gamma^i_{jk}=\frac{G^{il}}{2}(\partial_k G_{lj}+\partial_j G_{lk}-\partial_l G_{jk})
\label{Gamma}
\end{equation}
and indices $ijk$ along spatial directions. Here primes $'$ represent radial derivatives $\partial_r$.
This allows us to  define derivatives acting on spatial tensor fields, which are covariant under both spatial diffeomorphisms and radial shifts (\ref{shift_parameter}) as
\begin{equation}
\tilde{\nabla}_i w_j= \partial_i w_j + F_i\, w'_j-\tilde \Gamma^{l}_{ij}\,w_l.
\end{equation}
With these definitions we can write the so far unspecified  terms appearing in  (\ref{E:SigmaEOM})-(\ref{E:AEOM})
\begin{subequations}
\begin{align}
Q_\Sigma[\hat{g}]&=\frac{\text{tr}(g'^2)}{12}\\
P_{F}[\hat{g},\,\Sigma]_i^j&=-(G')_i^j+3\,\frac{\Sigma'}{\Sigma}\delta_i^j\\
Q_F[\hat{g},\Sigma]_i^j&= -(G'')_i^j+(G'^2)_i^j+\text{tr}(G''-\frac{G'^2}{2})\delta_i^j\\
S_F[\hat{g},\,\Sigma]_i&=\nabla_k (G')^{k}_i-6\,\nabla_i \frac{\Sigma'}{\Sigma}\\
 Q_{d_+\Sigma}[\Sigma]&=2 \, \frac{\Sigma'}{\Sigma}\\
 S_{d_+\Sigma}[\hat{g},\,\Sigma,\,F]&=\frac{\Sigma}{6}(\tilde{R}-12-\tilde \nabla F'-\frac{F'^2}{2})\\
  Q_{d_+\hat g}[\hat g, \Sigma]^{kl}_{ij}&=-(G')_{(i}^k\delta^l_{j)}+\frac{7}{2}\frac{\Sigma'}{\Sigma}\delta^k_{(i}\delta^l_{j)}\\
    \bar{S}_{d+\hat g}[\hat{g},\,\Sigma,\,F,\,d_+\Sigma]_{ij}&=\frac{1}{\Sigma^2}\Big(-\frac{3}{2} G'_{ij}\frac{d_+ \Sigma}{\Sigma}+\tilde{R}_{ij}-\tilde \nabla_j F'_i-\frac{F'_i\, F'_j}{2} \Big)\\
    S_{d+\hat g}[\hat{g},\,\Sigma,\,F,\,d_+\Sigma]_{ij}&=(\bar{S}_{d+\hat g})_{(ij)}-\text{tr}(\bar S_{d+\hat g})\frac{G_{ij}}{3}
\end{align} 
where $\tilde R_{ij}$ is the radial diffeomorphism covariant, spatial Ricci tensor.
\end{subequations}
\section{Decoupling of transverse derivative expanded Einstein equations}
\label{decoupling}
In the following we  are going to  show that the differential equations of the functions belonging to set $S_1$ in (\ref{set1}) can be decoupled from  the equations of the functions belonging to set $S_0$ in (\ref{set2}) through first order in transverse derivatives. Furthermore the differential equations of functions in $S_0$ do not depend on transverse derivatives of the zeroth order metric.\\ \indent 
Let the superscript $^{(1)}$ denote the first  order transverse derivatives correction to zeroth order in transverse derivatives solutions which we denote by the superscript $^{(0)}$. Then the following statements hold: 
\begin{enumerate}
\item \label{st1} For solutions at zeroth order in transverse derivatives we have $(\hat g_{ij})^{(0)}_{i \neq j}=0$ as well as $(F_x)^{(0)}=0$, $(F_y)^{(0)}=0$.
\item \label{st2}  The trace $\big(\text{tr}(g'^2)\big)^{(1)}$ does not depend on $(\hat g_{ij})^{(1)}_{i \neq j}$ or $(\hat g'_{ij})^{(1)}_{i \neq j}$ , since $(\hat g_{ij})^{(0)}_{i \neq j}=0$.
\item \label{st3}  $((G')^i_j)^{(1)}_{i\neq j}$ does not depend on  $(\hat g_{ij})^{(1)}_{i = j}$, $(\hat g'_{ij})^{(1)}_{i = j}$, $(\Sigma)^{(1)}$, $(\Sigma')^{(1)}$ and $((G')^i_j)^{(1)}_{i= j}$ does not depend on  $(\hat g_{ij})^{(1)}_{i \neq j}$ or $(\hat g'_{ij})^{(1)}_{i \neq j}$, again because  $(\hat g_{ij})^{(0)}_{i \neq j}=0$. 
\item \label{st4}  Similarly $((G'')^i_j)^{(1)}_{i\neq j}$, $((G'^2)^i_j)^{(1)}_{i\neq j}$  do not depend on  $(\hat g_{ij})^{(1)}_{i = j}$ and its derivatives nor on  $(\Sigma)^{(1)}$ and its derivatives. Also $((G'')^i_j)^{(1)}_{i= j}$, $((G'^2)^i_j)^{(1)}_{i= j}$ do not depend on  $(\hat g_{ij})^{(1)}_{i \neq j}$ or its derivatives.
\item \label{st5}  Again by only using that  $(\hat g_{ij})^{(0)}_{i \neq j}=0$, it is easy to show that  $ (\nabla_k (G')^k_{z})^{(1)}$ does not depend on transverse derivatives of zeroth order functions nor on  $(\hat g_{ij})^{(1)}_{i \neq j}$ and its derivatives. And  $ (\nabla_k (G')^k_{x,y})^{(1)}$ does not depend on $(\hat g_{ij})^{(1)}_{i = j}$,  $(\Sigma)^{(1)}$ and their derivatives.
\item \label{st6}  After a slightly more tedious calculation it follows again from statement \ref{st1} that $(\tilde R_{ij})^{(1)}_{i \neq j}$ does not depend on $(\hat g_{ij})^{(1)}_{i = j}$,  $(\Sigma)^{(1)}$,  $(F_z)^{(1)}$ and their derivatives and  $(\tilde R_{ij})^{(1)}_{i =j}$ does not depend on transverse derivatives of zeroth order functions nor on  $(\hat g_{ij})^{(1)}_{i \neq j}$ nor on $F_x$, $F_y$. The same is obviously true for $\tilde \nabla_j F'_i$ and $F'_i\, F'_j$.
\item \label{st7}  Statement \ref{st6} implies  that $\tilde R = G^{ij} \tilde{R}_{ij}$, $\tilde \nabla F'$ and $F'^2$ do not depend on transverse derivatives of  zeroth order functions, nor on  $(\hat g_{ij})^{(1)}_{i \neq j}$ nor on $F_x$, $F_y$. 
\item \label{st8}  Statement \ref{st3}, \ref{st4} and \ref{st5} imply  that  $(P_{F}[\hat{g},\,\Sigma]_i^j)^{(1)}_{i=j}$,  $Q_{F}[\hat{g},\,\Sigma]_i^j)^{(1)}_{i=j}$ and $S_F[\hat{g},\,\Sigma]_z$ do not depend on  $(\hat g_{ij})^{(1)}_{i \neq j}$, nor its derivatives, nor on  transverse derivatives of zeroth order functions and  $(P_{F}[\hat{g},\,\Sigma]_i^j)^{(1)}_{i\neq j}$,  $Q_{F}[\hat{g},\,\Sigma]_i^j)^{(1)}_{i\neq j}$ and $S_F[\hat{g},\,\Sigma]_{x,y}$ do not depend on  $(\hat g_{ij})^{(1)}_{i = j}$ and its derivatives nor on  $(\Sigma)^{(1)}$ and its derivatives.
\item \label{st9}  Statement \ref{st7} implies that  $( S_{d_+\Sigma}[\hat{g},\,\Sigma,\,F])^{(1)}$  does  not depend on transverse derivatives of zeroth order functions, nor on  $(\hat g_{ij})^{(1)}_{i \neq j}$, nor on $F_x$, $F_y$.
\item \label{st10}  Statement \ref{st1} and $\big((d_+g)^{(1)}_{ij}\big)_{i \neq j} =0$ imply that $ (Q_{d_+\hat g}[\hat g, \Sigma]^{kl}_{ij} d_+\hat g_{kl})^{(1)}$ for $i \neq j$ does not depend on $(d_+ \hat g_{ij})_{i=j}^{(1)}$ nor on  $(\hat g_{ij})^{(1)}_{i = j}$, $(\hat g'_{ij})^{(1)}_{i = j}$, $(\Sigma)^{(1)}$, $(\Sigma')^{(1)}$ and for $i = j$ it does not depend on $(d_+ \hat g_{ij})_{i\neq j}^{(1)}$, $(\hat g_{ij})^{(1)}_{i \neq j}$, $(\hat g'_{ij})^{(1)}_{i \neq j}$.
\item \label{st11}  Statement \ref{st1}, \ref{st6} and \ref{st7} imply that $(S_{d+\hat g}[\hat{g},\,\Sigma,\,F,\,d_+\Sigma]_{ij})^{(1)}_{i \neq j}$ does not depend on $(\hat g_{ij})^{(1)}_{i = j}$,  $(\Sigma)^{(1)}$ and their derivatives and  $(S_{d+\hat g}[\hat{g},\,\Sigma,\,F,\,d_+\Sigma]_{ij})^{(1)}_{i = j}$ does not depend on transverse derivatives of zeroth order functions, nor on  $(\hat g_{ij})^{(1)}_{i \neq j}$, nor on $F_x$, $F_y$.
\item \label{st12}  From Eq. (\ref{time_der_energy_(flux)}) it is obvious that the time derivative of  $(a^4)^{(1)}$ and $(f^4_{z})^{(1)}$ do not depend on $(f^4_{x,y})^{(1)}$, its derivatives, nor on transverse derivatives of $\hat{g}^4_{ij}$, nor on $(\hat{g}^4_{ij})_{i \neq j}^{(1)}$ and that the time derivative of $(f^4_{x,y})^{(1)}$ does not depend on $(f^4_{z})^{(1)}$, nor on $(\hat{g}^4_{ij})_{i = j}^{(1)}$.
\item \label{st13}  Due to \ref{st1}, the time derivative of $(\hat g_{ij})^{1}_{i \neq j}$ does not depend on $A^{(1)}$, nor on $\partial_t \lambda^{(1)}$. In addition  $(\hat g_{ij})^{1}_{i \neq j}$ itself is not affected by first order shifts $\lambda^{(1)}$.
\item \label{st14}  From Eq. (\ref{Eq_lambda}) it is apparent that the first order correction to the shift function $\lambda^{(1)}$ does not depend on transverse derivatives of the  zeroth order metric, nor on $(F_x)^{(1)}$, $(F_y)^{(1)}$. 
\end{enumerate}
From statement (\ref{st8})-(\ref{st14}) both claims follow.

\end{appendices}

%----------------------------------------------------------------------------------------
%	WORK EXPERIENCE SECTION
%----------------------------------------------------------------------------------------
\FloatBarrier

\end{document}